\documentclass[usenatbib]{mn2e}

\usepackage{times} 
\usepackage{graphicx}
\usepackage{rotating}
\usepackage{psfig}
\usepackage{epstopdf}
\usepackage{subfigure}
\usepackage{lscape}
\usepackage{longtable}
\usepackage{wasysym}
\usepackage{psfig}
\usepackage{amssymb}

\newcommand{\lsol}{\hbox{$L_\odot$}}
\newcommand{\e}[1]{$10^{#1}$}
\newcommand{\ee}[1]{$\times 10^{#1}$}         
\newcommand{\kms}{~km\,s$^{-1}$} 
\newcommand{\cm}[1]{~cm$^{#1}$}
\newcommand{\ergs}{~erg\,s$^{-1}$\,cm$^{-2}$\,sr$^{-1}$}

\newcommand{\co}{$^{12}$CO}    
\newcommand{\tco}{$^{13}$CO}
\newcommand{\htco}{H$_2$CO}
\newcommand{\hcop}{HCO$^+$}
\newcommand{\oh}{OH(1720~MHz)\ }
\newcommand{\tmb}{$T_{mb}$}
\newcommand{\tkin}{$T_{kin}$}
\newcommand{\tex}{$T_{ex}$}
\newcommand{\nh}{$n({\rm H}_2)$}
\newcommand{\h}{H$_2$}

\newcommand{\hr}{H\,{\sc ii}}
\newcommand{\hone}{2.12~$\micron$ H$_2$ 1--0 S(1)\ }
\newcommand{\htwo}{2.25~$\micron$ H$_2$ 2--1 S(1)\ }

\newcommand{\mjb}{\,mJy\,beam$^{-1}$}
\newcommand{\candy}{G349.7+0.2}

\newcommand{\aap}{A\&A}
\newcommand{\aaps}{A\&AS}
\newcommand{\aj}{AJ}
\newcommand{\apj}{ApJ}
\newcommand{\apjl}{ApJL}
\newcommand{\apjs}{ApJS}
\newcommand{\mnras}{MNRAS}

\newcommand{\nat}{Nature}

\begin{document}

%%% NOTE - ATNF IS NOW CASS

\title[SNR G349.7+0.2 shocks in a molecular cloud]
{Multiwavelength observations of the supernova remnant G349.7+0.2  
interacting with a molecular cloud}
\author[Lazendic et\ al.]
         {J. S. Lazendic$^1$, M. Wardle$^2$, J. B. Whiteoak$^3$, 
 M. G. Burton$^4$ and A. J. Green$^5$  \\
 $^1$ School of Physics, Monash University Clayton, VIC 3800, Australia\\
 $^2$ Department of Physics, Macquarie University, NSW 2109, Australia\\
 $^3$ Australia Telescope National Facility, CSIRO, PO Box 76, Epping
NSW 1710, Australia \\
 $^4$ School of Physics, University of New South Wales, NSW 2052, Australia\\
 $^5$ Sydney Institute for Astronomy, School of Physics, University of Sydney, NSW 2006, Australia}

\date{Accepted for publication in MNRAS, 2010, vol. 409, p. 731}

%\pubyear{2001}

\maketitle

\begin{abstract}
We present molecular-line observations at millimetre, 
centimetre and infrared wavelengths of the region containing \oh masers in 
the supernova remnant (SNR) \candy, using the Australia 
Telescope (AT) Mopra antenna, the
Swedish-ESO Submillimeter Telescope, the AT Compact Array and the UNSW
 Infrared Fabry-Perot narrow-band filter installed on 
the Anglo Australian Telescope. Several
 molecular transitions were observed between 1.6 and 3~mm  
 to constrain the physical parameters of the molecular cloud interacting
 with the SNR and to investigate the effects of
 the SNR shock on the gas chemistry. We detected shock-excited 
near-infrared \h\ emission towards the
centre of the SNR, revealing highly clumped molecular
 gas and a good correlation with published mid-infrared images from 
the {\em Spitzer Space Telescope}. An excellent correlation between the \h\ clumps and \oh maser
 positions supports the shock excitation of the \oh maser
 emission. Furthermore, we detected OH absorption at 1665 and 1667 MHz
 which shows a good correlation with the shocked \h\ emission and the
 masers. We found maser emission at 1665~MHz near the \oh masers
in this SNR, which is found to be associated with a GLIMPSE source SSTGLMC~G349.7294+00.1747.
We also detected 1665 and 1667~MHz OH masers, and weak 4.8~GHz \htco\ absorption towards the
 ultra-compact \hr\ region IRAS~17147--3725 located to the southeast of the SNR.
 We found no 4.7- or 6-GHz excited-state OH masers or 6-GHz CH$_3$OH
maser towards either the SNR or the \hr\ region.

\end{abstract}


\section{Introduction}

The young, massive stars which are progenitors of supernova remnants
(SNRs)  live only a relatively short time (a few million years) and
do not move far from the molecular clouds from which they are born.
Hence it is expected that SNRs formed from such massive stars should
be found in close proximity to their parent molecular clouds.  The shocks
driven by SNRs into dense molecular clouds  compress, accelerate and
heat the gas. They can  partially or completely disrupt the clouds and
may initiate star formation by triggering further gravitational
collapse following the cooling of the compressed gas. The shocks also
provide energy which can potentially excite higher molecular
transitions and activate chemical reactions forbidden in cold
molecular clouds, changing the chemical abundances in the cloud. Thus,
observations of molecular clouds which have interacted with a supernova
blast wave can provide important information about the physical and
chemical processes associated with the shocks.

For many years, the best unambiguous case of an SNR-molecular cloud 
interaction was the object IC~443, where broadened molecular
lines \citep*[see][and references therein]{denoyer79,vand93} and shocked molecular 
 hydrogen \citep{burton88} were detected. However, recent detections of 
\oh maser emission near the boundary regions of $\sim 10\%$ of the Galactic 
SNRs \citep{frail94,frail96,yz96-1,green97,kor98}  have become  
 signposts of the SNR interaction with molecular
clouds.  This maser emission, usually unaccompanied by maser emission
in the other three ground-state transitions, is excited from
 OH collisions with \h\ molecules in a gas with 
temperatures between 50--125 K, a  
density of $\sim$ \e{5}\cm{-3} and an OH column density between
\e{15} and \e{17}\cm{-2} \citep*{elitzur76,lock99}. These conditions can
be found in cooling gas behind a non-dissociative C shock, 
irradiated by the X-ray flux from the SNR interior \citep{wardle99}.
 Hence, SNRs associated with this maser emission are good candidates for
 studies of shock phenomena in molecular clouds. Follow-up molecular-line 
 observations of these SNRs have confirmed the presence of shocked molecular gas
towards the \oh maser locations \citep[eg][]{frail98,reach99,reach00,lazendic02,reach02,reach05,lazendic04,reach06}. Furthermore, 
new observations with improved sensitivity raised the fraction of maser-emitting SNRs to 15\%, implying that this fraction might become even higher with deeper searches for \oh\ masers \citep{hewitt09b}.

 One of the SNRs associated with \oh maser emission is \candy. This  
 is the third brightest Galactic SNR at radio wavelengths, after
Cas~A  and the Crab Nebula \citep{shav85}. It is classified as a
shell-type remnant from the radio continuum image, which has a circular shape
 ($\sim 2.5$~arcmin in diameter) with a bright periphery, and a typical
spectral index of $-0.5$. However, it has a southern enhancement that is 
different from the morphology of typical
shell SNRs. Five \oh masers have been detected along the bright
emission ridge of the   SNR \citep{frail96} with radial LSR (local
standard of rest) velocities ranging from 14.3 to 16.9\kms.  Their
 properties are listed in Table~\ref{tab-masers}. A 
magnetic field with a strength of
$0.35\pm0.05$~mG has been measured towards the brightest of the
masers \citep{brogan00}. The distance to the SNR is estimated to be 18.3$\pm$4.6~kpc  
from H\,{\sc i} absorption measurements  \citep{casw75}, and
 the maser velocities are consistent with this value 
\citep[$\sim 22$~kpc;][]{frail96}.  The SNR is predicted to be $\sim 3000$\,yrs old and its 
X-ray morphology is strikingly similar to the radio morphology \citep{lazendic05}, suggestive 
of expansion into a medium with a large density gradient, as has been  
 observed in the \co\ 1--0 line by \citet{estela2001}, with a resolution of 54~arcsec. 
 \citet{dubner04} observed \co 1--0, 2--1 and 3--2 lines towards the central 
part of the SNR and the location of OH masers, and found the line ratios indicative of 
 shocked molecular gas.

To investigate the effects of SNR shocks on interstellar molecular
 gas and to test the models for the generation of \oh emission via shocks, we 
carried out molecular-line observations in both radio and infrared
 bands. We used observations of molecular  hydrogen to trace recently
 shocked gas identified by the presence of the masers. Molecular lines
 at millimetre wavelengths were used to probe the structure, dynamics and
 composition of the  molecular gas in which masers are created. OH
 line observations at centimetre wavelengths (1665 and 1667~MHz) were used to 
 derive OH column densities and to test the model  for \oh maser production. 
We also searched for excited-state OH maser emission at 
4.7 GHz and 6.0 GHz, CH$_3$OH maser emission at 6 GHz, 
and \htco\ absorption at 4.8~GHz. We
 summarize the observations in Section~\ref{sec-candy2} and present 
the results in Section~\ref{sec-candy3}.  The results are discussed in
 Section~\ref{sec-candy4}, and our conclusions are given in
 Section~\ref{sec-candy5}.

\section{Observations}
\label{sec-candy2}

\subsection{Molecular-line observations at mm wavelengths}

To obtain the extent of the molecular cloud associated
with \candy\ we used the 22~m
Australia  Telescope Mopra antenna during October 1998.
Observations of the 3-mm 1--0 transition of \tco\ were undertaken on
 a $7\times7$ point grid centred at  ${\rm
RA}(2000)=17^{\rm h}~18^{\rm m}~00^{\rm s}$, ${\rm
Dec.}(2000)=-37\degr~26\arcmin~10\arcsec$, with a 30~arcsec grid and
60~second integration per position. 
For observations of other molecular species (\co, CS, \hcop, HCN,
\htco, SiO and SO) we used the 15~m Swedish-ESO Submillimetre Telescope 
(SEST) telescope in February and June 1999. The observed
transitions, their frequencies and corresponding beamwidths are given
in Table~\ref{tab-molecules}.  Most of the transitions
were observed over a  $2\times2.5$ arcmin$^2$ region centred at  ${\rm
RA}(2000)=17^{\rm h}~18^{\rm m}~00^{\rm s}$, ${\rm
Dec.}(2000)=-37\degr~26\arcmin~30\arcsec$, with a 24~arcsec grid
spacing and
60 second integration per position. Due to the
overall weak emission of the CS, \htco, SO and SiO lower energy transitions, higher 
transitions for these species were observed only at the
maser positions in a five-point cross grid with 20~arcsec offset. 
For all transitions a position-switching observing mode 
 was used with a reference position at ${\rm
RA}(2000)=17^{\rm h}~13^{\rm m}~23^{\rm s}$, ${\rm
Dec.}(2000)=-37\degr~03\arcmin~30\arcsec$.

When the \tco\  observations at 110~GHz were carried out with the Mopra 
telescope, the effective diameter of the dish was 15~m with a 
corresponding beamwidth of about 43~arcsec, slightly smaller than the SEST
beamwidth of about 45~arcsec for the 115-GHz \co\ observations. For
the Mopra observations, the attenuation due
to atmospheric absorption was corrected using measurements of
a black body paddle at ambient temperature 
\citep[see][for more details]{hunt97}.  
The corrected intensities were then matched
to the main-beam temperature scale of the SEST by observing the
molecular cloud in the direction of the Ori~A SiO maser and scaling
the observed intensities to their SEST counterparts. The final
intensity calibration is believed to be accurate to within $15\%$. The SEST observations 
were similarly corrected for atmospheric attenuation during the observations, and 
then corrected for the main-beam efficiency (0.74, 0.70, 0.67 and 0.45 at 
85--100~GHz, 100--115~GHz, 130--150~GHz and 220--265~GHz respectively). 
For both telescopes the pointing was corrected approximately every 2  hours by 
observations of the 86~GHz SiO masers of AH\,Sco and W\,Hyd.

\subsection{Near-infrared observations}

The NIR observations were carried out in June 1999 using the
University of New South Wales Infrared Fabry-Perot narrow-band filter
\citep[UNSWIRF;][]{ryder98}, in conjunction with
the Infrared Imager and Spectrometer  \citep[IRIS;][]{allen93} on the 3.9~m
Anglo-Australian Telescope (AAT). A pixel size of
0.77~arcsec resulted in  a circular image of 100~arcsec in
diameter. Observations were obtained in the \hone transition, centered
at ${\rm RA}(2000)=17^{\rm h}~17^{\rm m} 58^{\rm s}$, 
${\rm Dec.}(2000)=-37\degr~26\arcmin~27\arcsec$, 
and  the \htwo transition, with a similar position.
%RA(2000) $17^{\rm h}~18^{\rm m} 00^{\rm s}$ Dec.(2000)
%$-37\degr~26\arcmin~26\arcsec$. 
Each observation consisted of a set of  five Fabry-Perot frames, 
equally spaced by 40\kms, centred on the average maser velocity. 
The velocity resolution was $\approx$
75\kms.  The  integration time was 180 seconds per frame for all
observations. The data were  reduced using modified routines in the
{\tt IRAF} software package  \citep{ryder98}. An off-line frame,
with velocity offset by $-400$\kms\ from the first  on-source frame, was
taken for subsequent continuum subtraction. The  intensity was
calibrated using additional observations of the standard star BS
6441.  Final data cubes were fitted with the instrumental Lorentzian
profile to  determine the line flux for the \h\ emission across the
field. The resultant  uncertainty in the line flux is  less than $30\%$. It should be 
noted that the images are not true velocity channel maps, because each pixel in a frame has a slightly different line centre velocity.  Both the central wavelength
and the spectral resolution vary between pixels (in a reproducible
manner)  by up to $\approx20$\kms. Calibration was achieved using an arc line, 
to give line centre maps.

To establish the spatial coordinates for the \h\ images, we carried out observations
with the Cryogenic  Array  Spectrometer Imager \citep[CASPIR;][]{mcgr94}
 camera installed on  the 2.3~m telescope at the Siding Spring
Observatory in June 1999. The observations  were  
undertaken using standard filters: $J$
(1.15--1.4 $\micron$),  $H$ (1.55--1.8 $\micron$) and $K$ (2.0--2.4
$\micron$). The selected pixel size of 0.5 arcsec yielded
 individual frames with a size of $\approx 2$ arcmin. A total of nine frames,
with  20 arcsec spatial overlap, was obtained with each filter. The exposure time
was 180  seconds for an individual frame.  For each filter, 
bias and dark  frames were taken, as well as frames of the dome lamp light
for flat-fielding. The  data were reduced using modified routines in
{\tt IRAF}. No extended emission  was detected and median sky frames were
made by combining six neighboring source frames. To make the final
image, the nine observed frames per filter were  combined in a mosaic
covering a region of $\approx 5 \times 5$~arcmin$^2$. For the 
 astrometry, stars in the $J$-band mosaic were matched with Digital Sky 
Survey (DSS)  stars. Stars which were also detectable in the
$K$-band mosaic were then used as reference stars for  astrometry of the
UNSWIRF frames. The absolute positions have a maximum uncertainty
better than 1~arcsec.

\subsection{Molecular-line observations at cm wavelengths}

We used the  Australia Telescope Compact Array (ATCA),
which consists of six 22~m antennas, to observe
ground- and excited-state transitions of OH at 1.6, 4.7 and
6.0~GHz, the 4.8~GHz transition of \htco\ and the 6.6~GHz transition
of CH$_3$OH. The  observational parameters are listed in
Table~\ref{tab-atca} and the observed molecular transitions in
Table~\ref{tab-atca-trans}.  For some observations the correlator 
configuration also allowed a wideband setting (128~MHz band over 32 channels)
 for observations of continuum emission, in addition to the 
narrow spectral-line mode (see Table 3). 
The calibration source PKS~1934--638 was observed at each frequency to
provide primary flux density and bandpass calibration.
An observing cycle was used in which target source integrations
 were bracketed by short  observations of the phase calibrator
PKS~1718--649 used for calculation of the complex antenna gains.  For
the OH ground-state transitions the observing band was centred at 1666~MHz to 
include emission from both `main' lines at 1665 and 1667~MHz. The OH 
excited-state observations consisted of two or three contiguous 10-minute
integrations on-source with the observing band set at the different
transition frequencies for each integration.   
Line-free channels were used for subtraction of  the
continuum emission from all channels to form spectral-line data
cubes and, in some cases, to provide continuum images. The rms noise values for the 
final spectral-line cubes are listed in Table~\ref{tab-atca}.  

A continuum image at 18~cm was produced from two datasets 
at  1666~MHz formed by averaging about 350 of the 1024 channels, which were
line-free. The continuum  image at 6~cm   was
constructed by combining data at 6035~MHz and 4829~MHz. For the
 6035~MHz observations the 128~MHz  wideband configuration was used. However,
 for the 4829~MHz data, no \htco\ was detected and hence, the central 1800 
 channels of the 4~MHz spectral band were able to be used. 
Because of different pointing centres and central frequencies for the two datasets, the 
images were produced using  mosaicing and multi-frequency synthesis techniques 
from the {\tt MIRIAD} software package \citep{sault97}.

%%%%%%%%%%%%%%%%%%%%%%%%%%%%%%%%%%%%%%%%%%%%%%%%

\section{Results and analysis}
\label{sec-candy3}

\subsection{SNR morphology}
\label{sec-morphology}

\begin{figure*}
\centering
\includegraphics[height=8cm]{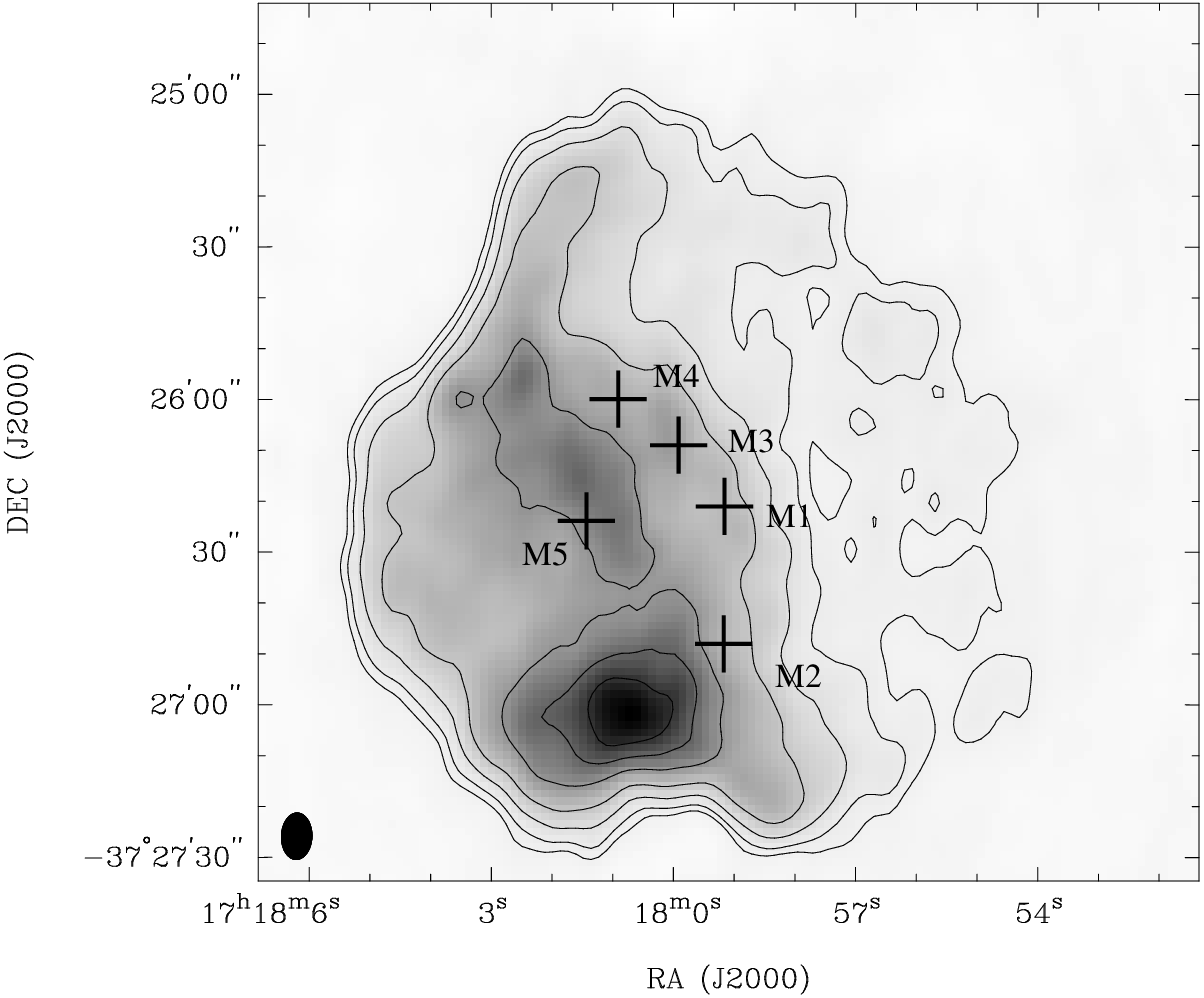}\\
\includegraphics[height=8cm]{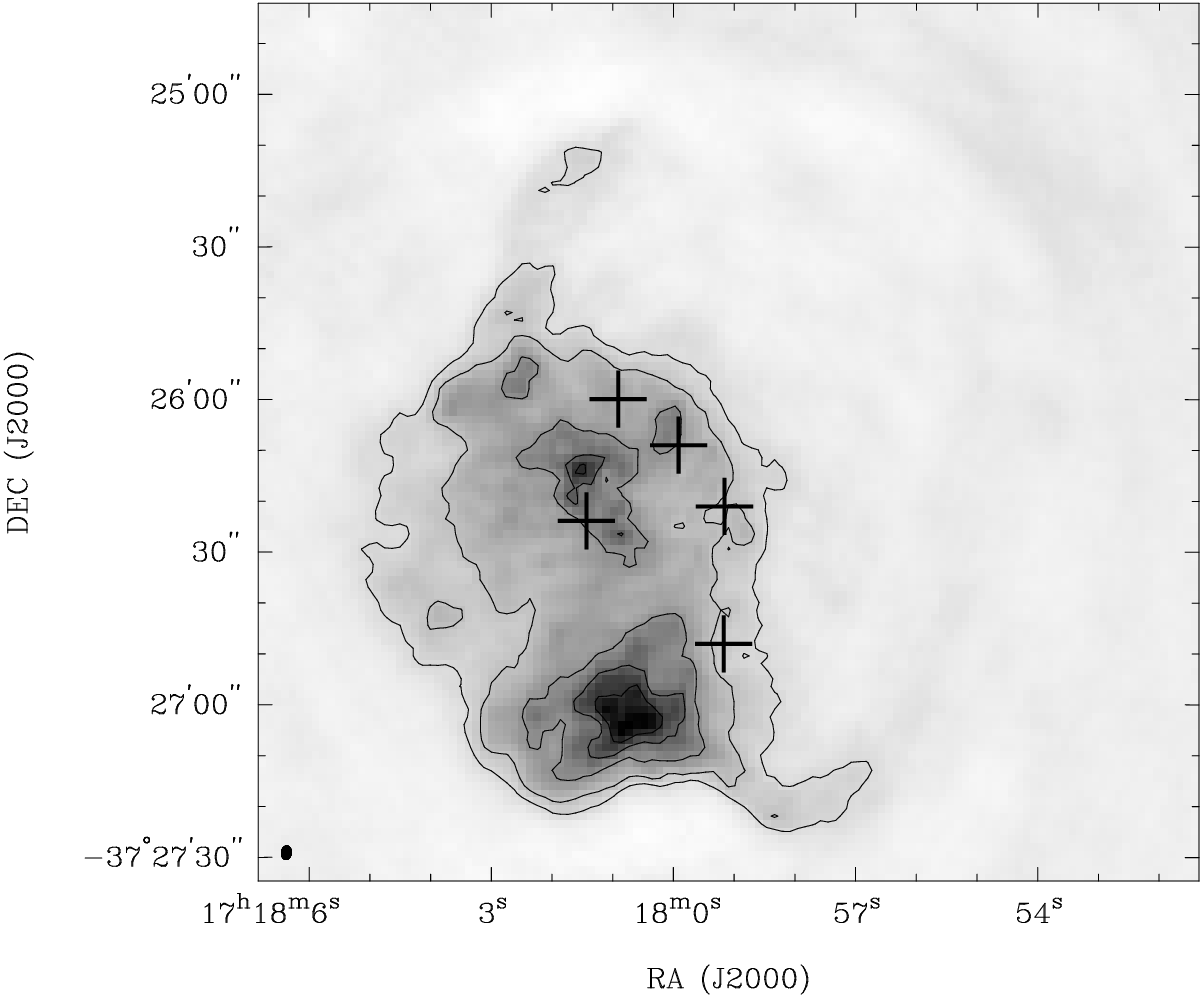}
\caption{The ATCA continuum images of \candy\ at 18~cm ({\it top}) with an rms noise of 1.5~mJy/beam, and 6~cm ({\it bottom}) with an rms noise of 0.2~mJy/beam, shown in greyscale and contours. Contour levels are: (18~cm): 12, 17, 34, 68, 136, 204, 272 \mjb\ and (6~cm): 1.6,
3.2, 6.4, 10, 13 \mjb. The crosses mark the \oh maser positions. The
synthesised beam is shown in the lower left corner.}
\label{fig-radio}
\end{figure*}

In Figure~\ref{fig-radio} we show the continuum images of \candy\ at 18
and  6~cm.  The 18~cm image has a FWHM synthesized beam of 9\farcs 0
$\times$ 5\farcs 8  (P.A.  = $-$1\fdg 0),  an rms noise of
1.5\mjb and is similar to existing radio images  of the region
\citep{frail96,brogan00}.  The 6~cm image has a synthesised beam
of  2\farcs 5 $\times$ 1\farcs 8  (P.A. = $-$5\fdg 7) and an rms noise
of 0.2\mjb. Only the bright eastern part of the SNR was detected at
6~cm. The emission to the west is too weak to be detected
 at the higher frequency with the available sensitivity and resolution.
 A consequence of the shortest baselines of the ATCA  configurations used for
 this work is that  the 6~cm observations are only sensitive to structures $\le 2$ arcmin 
 in extent, while at 18~cm structures larger in scale than 5--8 arcmin will be resolved out.  
 The integrated flux densities are 15.5 and 6.6~Jy at 18 and 6~cm,  respectively. These 
 values are lower limits because the observations are of an extended source, subject 
 to some flux underestimation because of the limitation of an interferometer in detecting 
 large spatial scales. Nevertheless, we can estimate a mean spectral index of $-0.7\pm0.3$ 
 between these two  wavelengths, which is in agreement with the previously 
derived value of $-0.5$ \citep{shav85}. 

\begin{figure*}
\centering
\includegraphics[height=10cm]{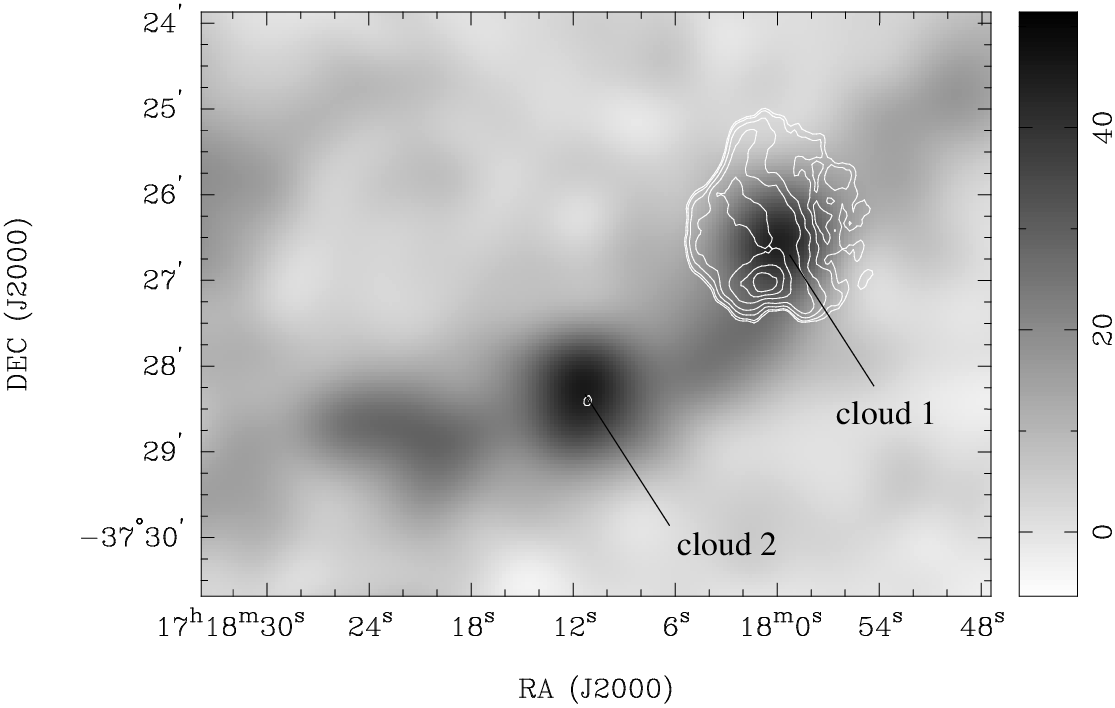}
\caption{Greyscale image of the 
\co\ 1--0 emission obtained with the 12-m NRAO (National
Radio Astronomy Observatory) antenna, integrated between +14.5 and
+20.5\kms\ and convolved to a 60 $\times$ 60 arcsec$^2$ beam 
 from \citet{estela2001}. The contours represent the 18 cm continuum
emission from \candy\ with the levels same as in
Figure~\ref{fig-radio}. Cloud~1 is coincident with the SNR \candy, and
cloud~2 is coincident with an UC \hr\ region IRAS~17147--3725 located to the 
southeast of the SNR, represented by a single radio contour.}
\label{fig-estela}
\end{figure*}

The influence of the interstellar medium on the radio morphology of \candy\ has  been 
studied by \citet{estela2001}. They mapped the \co\ 1--0 distribution around the SNR over 
an area of $38 \times 38$~arcmin$^2$ with a resolution of 54~arcsec. Figure~\ref{fig-estela} 
shows their \co\ map, integrated between +10 and +20\kms\ , to demonstrate that the 
molecular cloud associated with the SNR (`cloud 1') is part of a complex encompassing
another cloud south-east of the SNR (`cloud 2'), which is
coincident with the low-brightness radio continuum source designated
as ultra-compact (UC) H\,{\sc ii} region IRAS~17147--3725
\citep{wood89,bronfman96}. Reynoso \& Mangum (2001) suggested that the SNR 
morphology is influenced by the density gradient in the surrounding
molecular gas, with the bright ridge encountering denser gas, and
the fainter section of the SNR shell expanding into a lower density gas. The 
X-ray morphology shown in images from  the {\em Chandra X-ray Observatory} supports 
this inference  \citep{lazendic05}.

\subsection{Millimetre-line emission} 

\begin{figure*} 
\centering
\includegraphics[height=7cm]{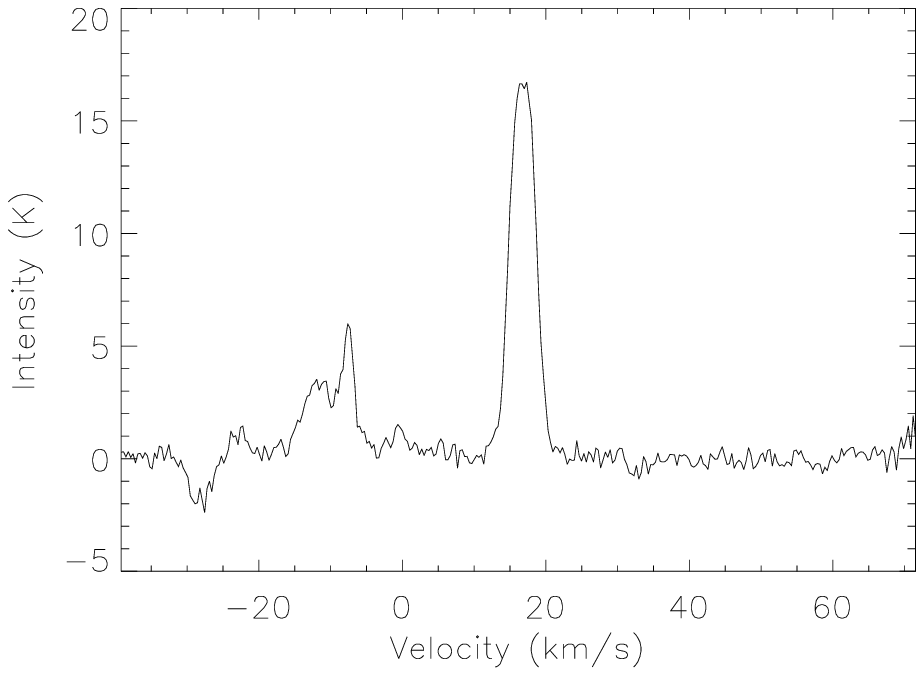}
\caption{\co\ 1--0 spectrum (intensity scale is in antenna temperature ($T_a$) units) towards the SNR \candy\ in the range $-40$ to +80\kms. Molecular clouds are present at $-10$, +6 and +16\kms. The absorption around $-$30\kms indicates that spectra used for baseline subtraction had emission at that velocity.}
\label{fig-features}
\end{figure*}

\begin{figure*} 
\centering
\includegraphics[height=3.5cm]{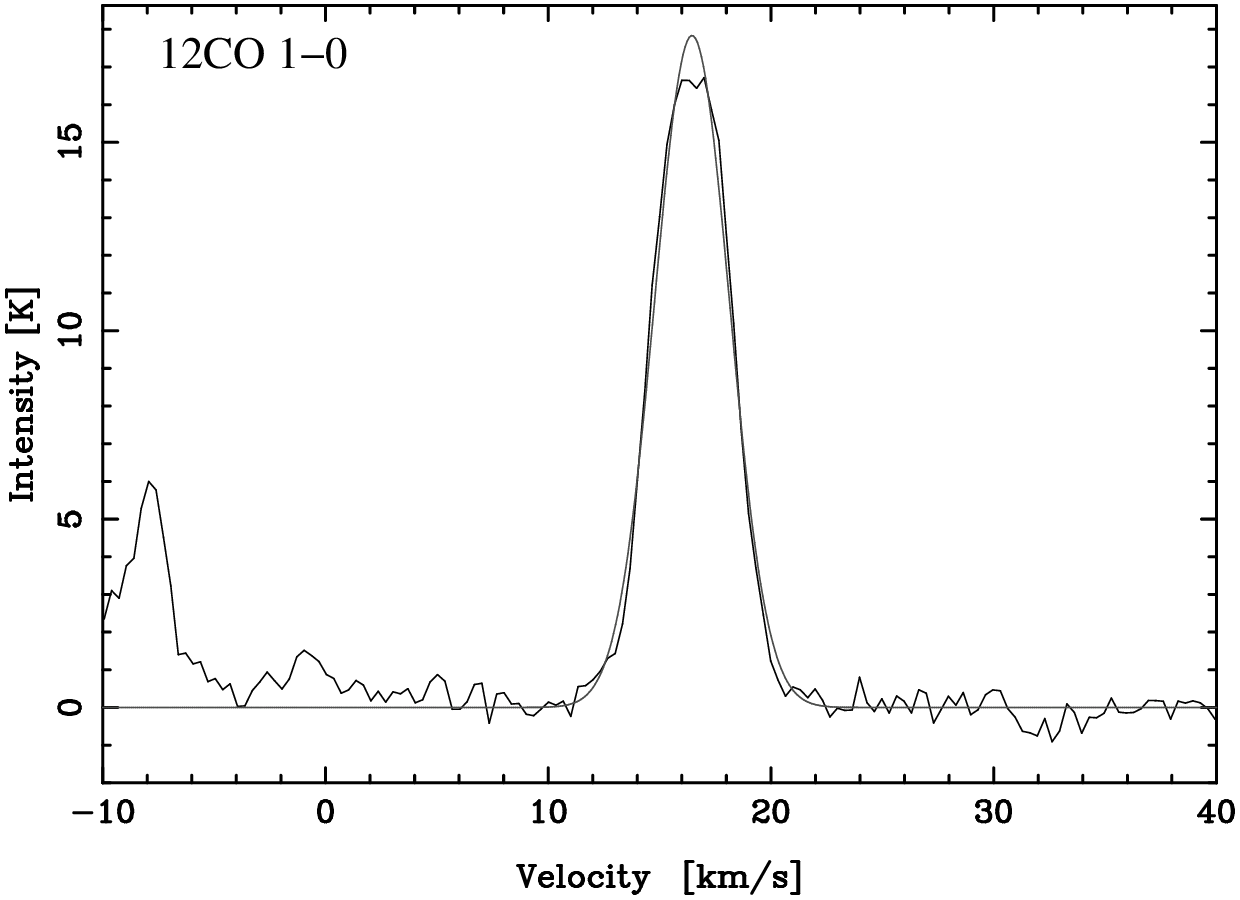}
\includegraphics[height=3.5cm]{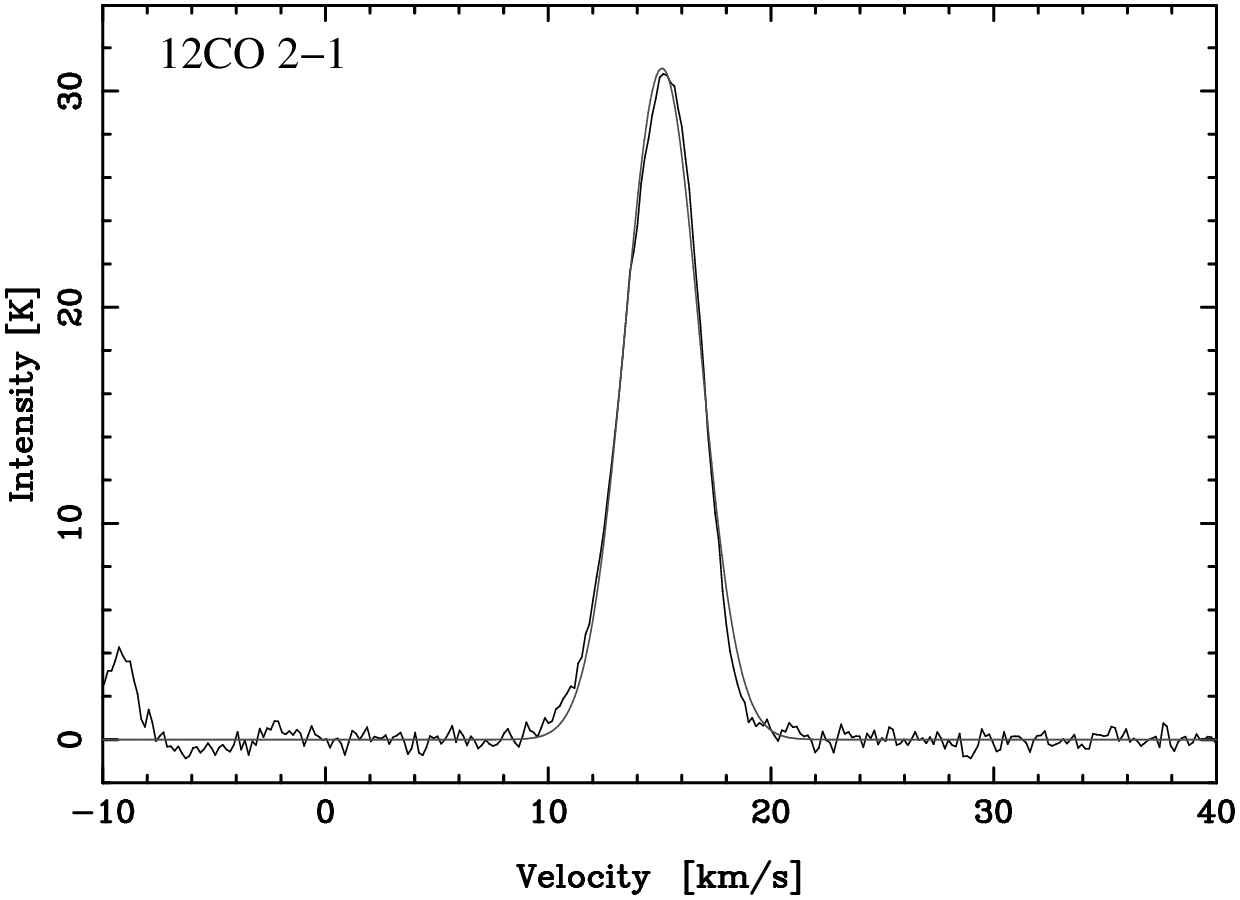}
\includegraphics[height=3.5cm]{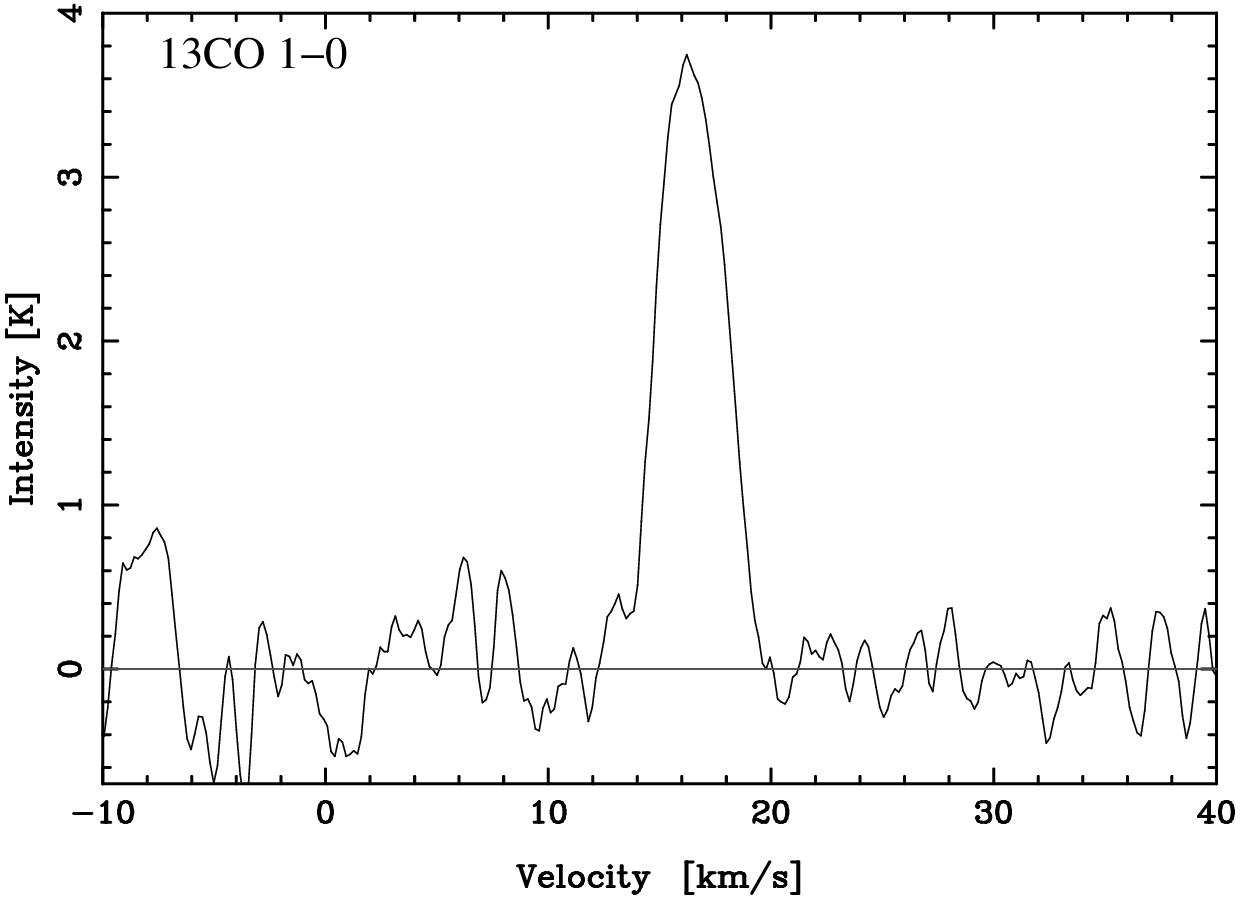}\\
\includegraphics[height=3.5cm]{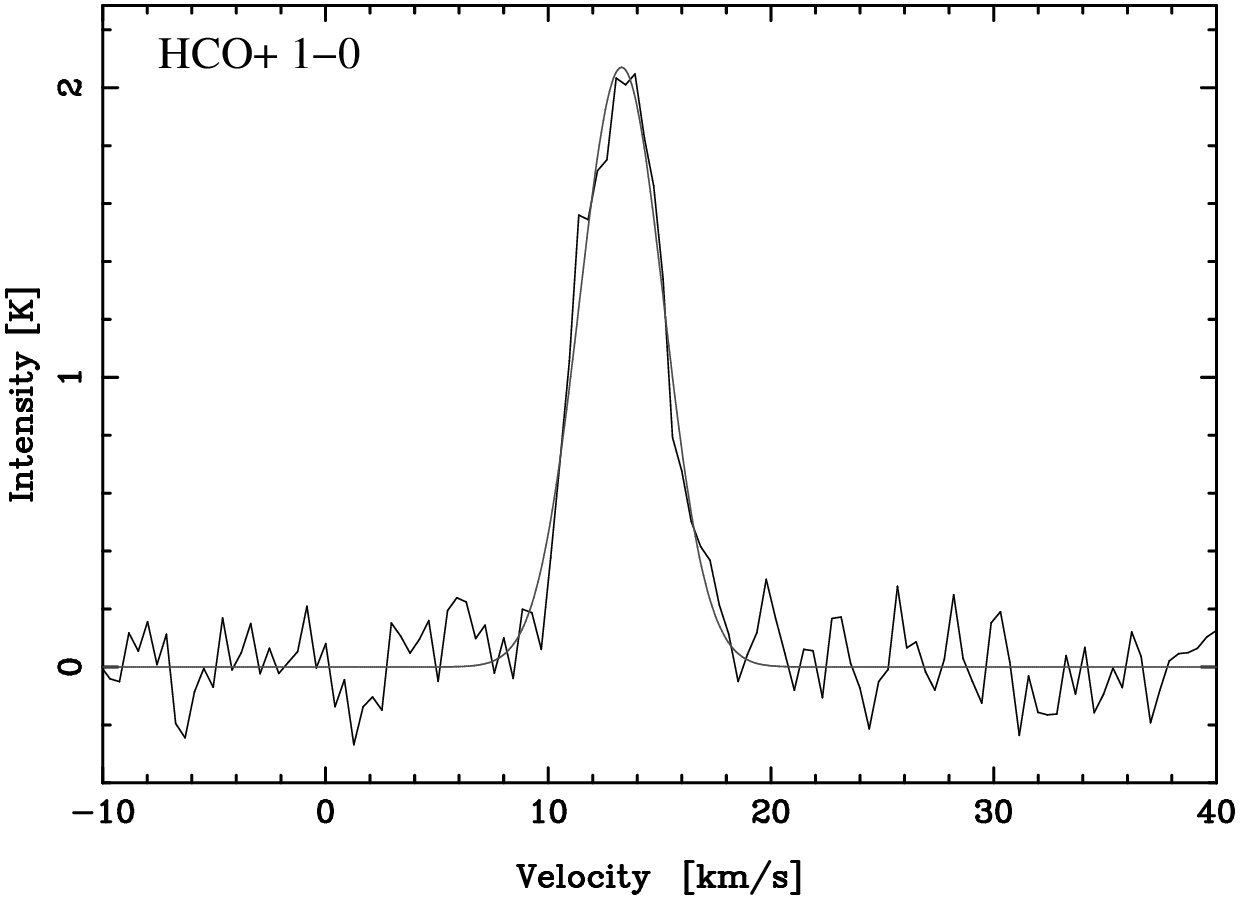}
\includegraphics[height=3.5cm]{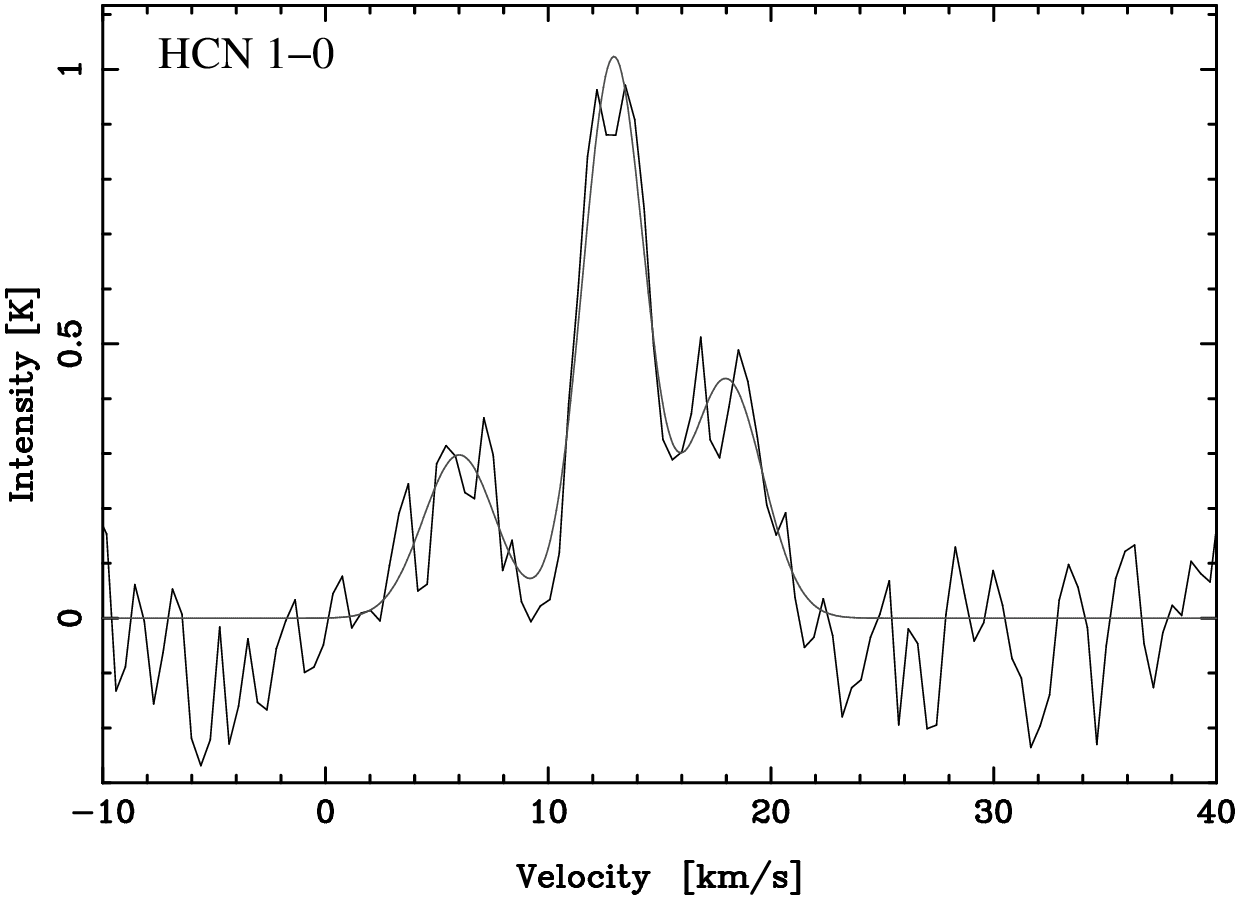}
\includegraphics[height=3.5cm]{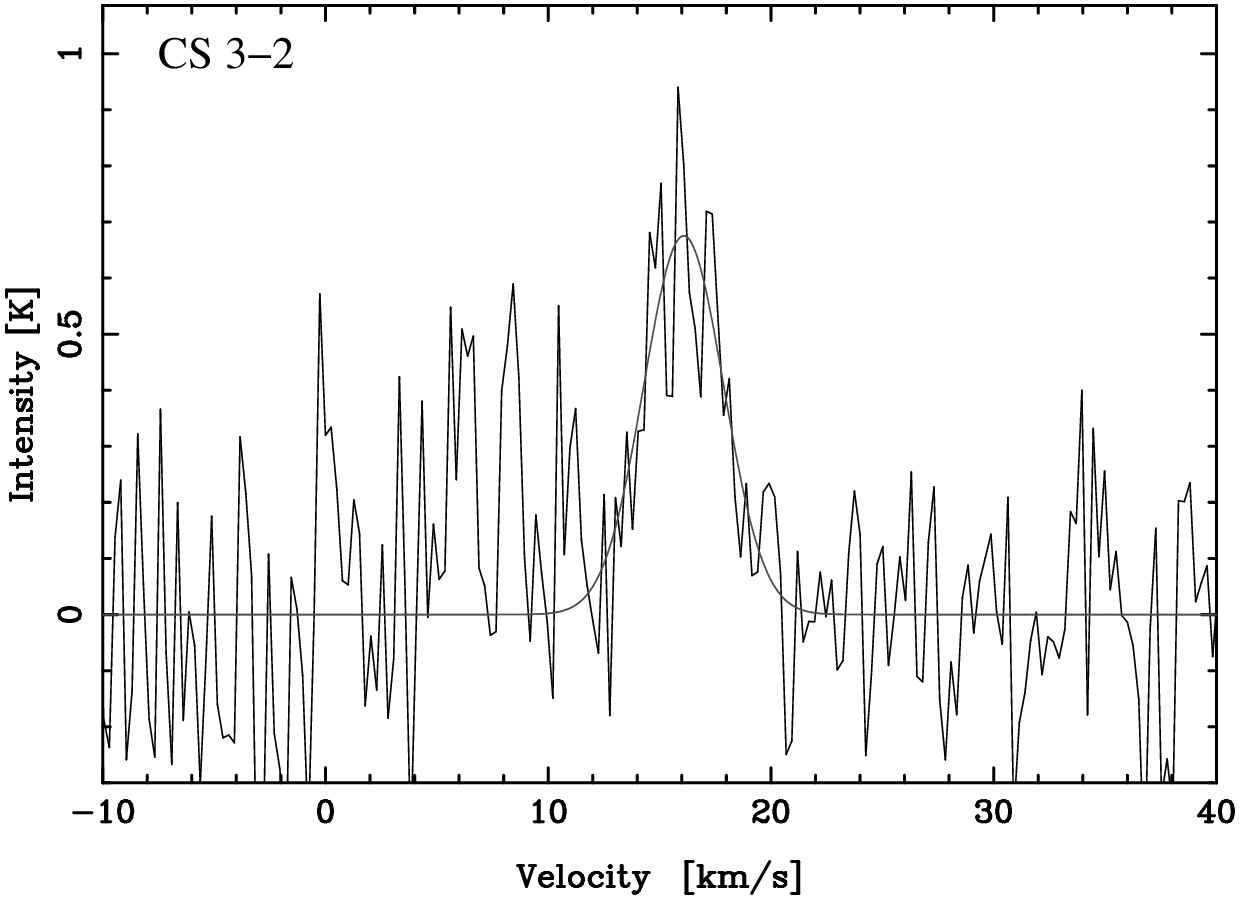}\\
\includegraphics[height=3.5cm]{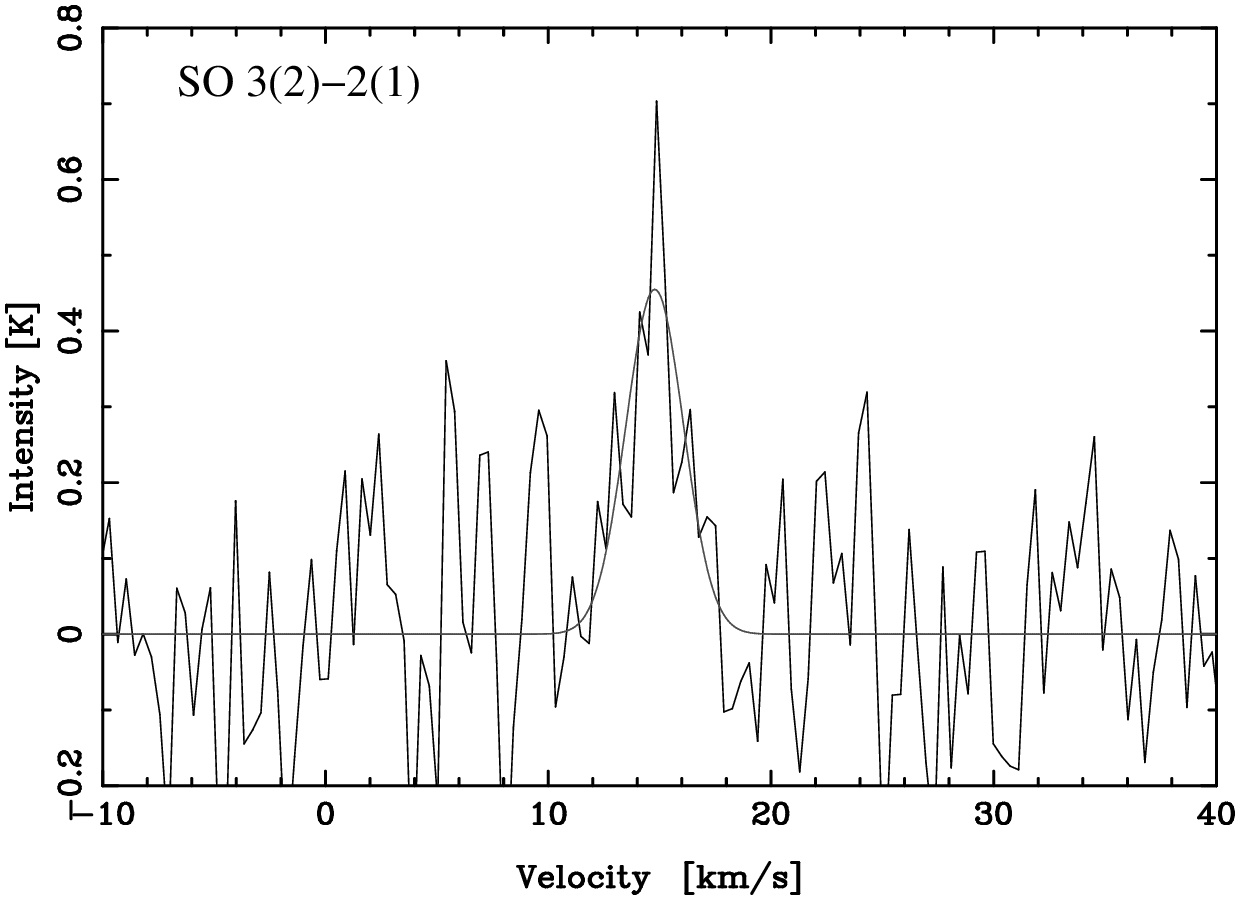}
\includegraphics[height=3.5cm]{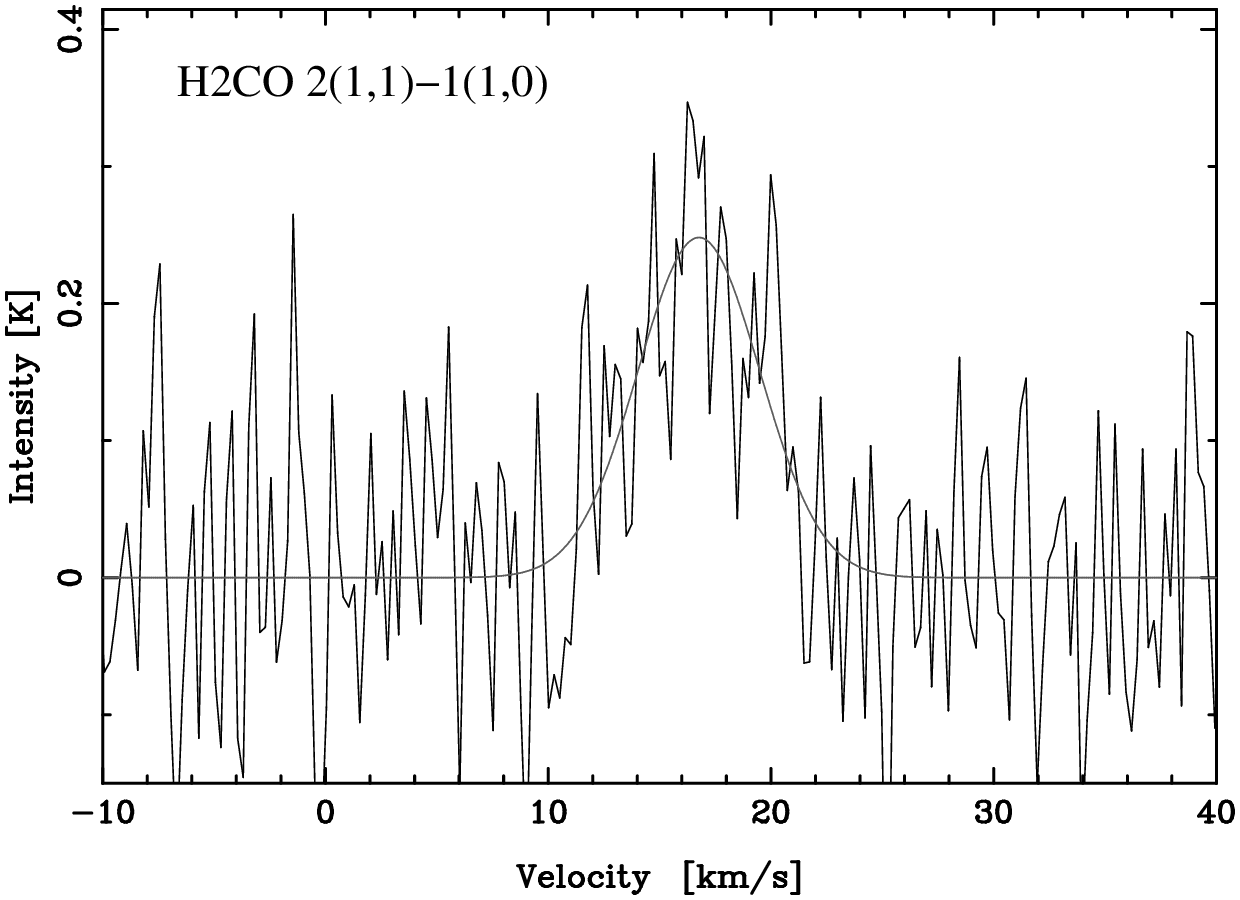}
\includegraphics[height=3.5cm]{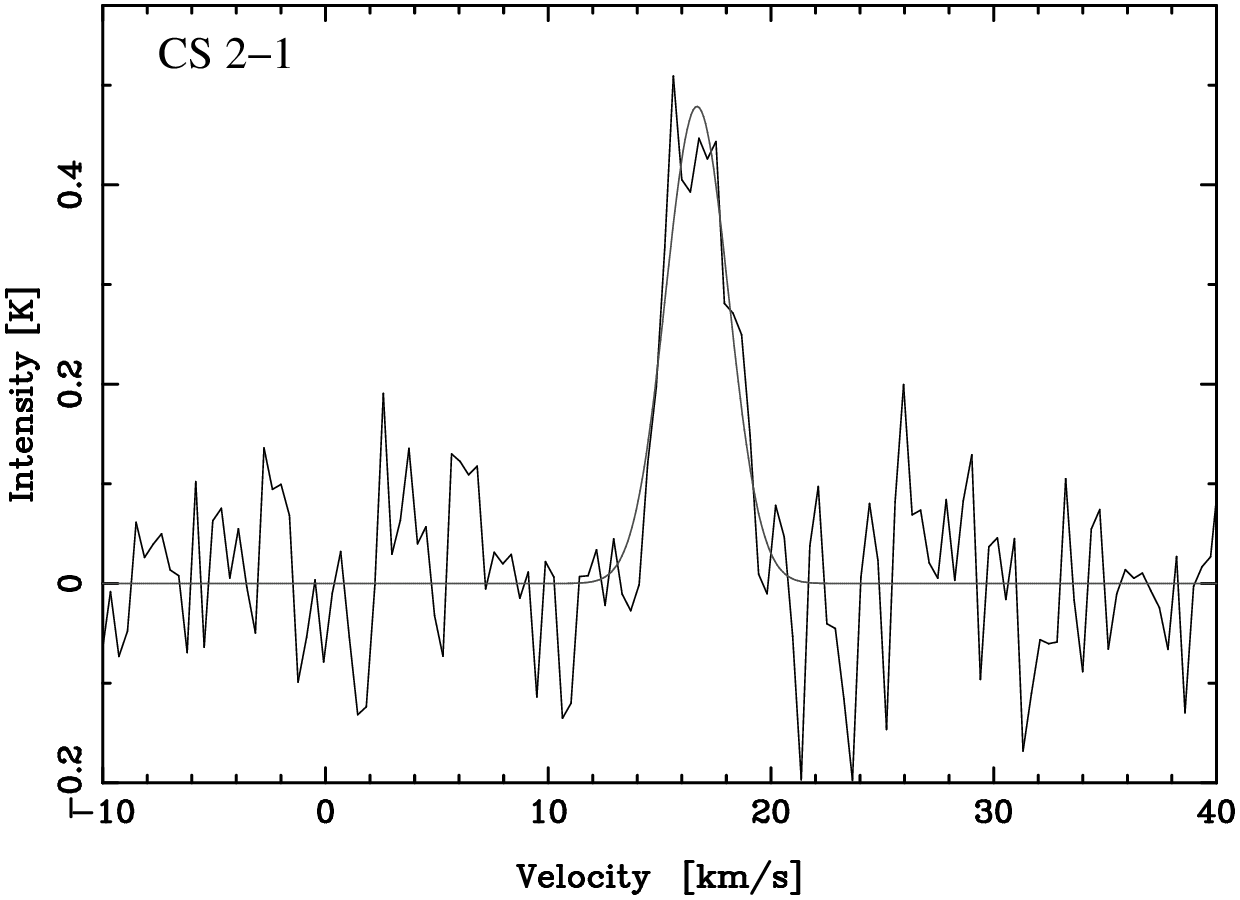}
\caption{The observed millimetre-line profiles of the +16\kms\
molecular cloud towards \candy, overlaid with the Gaussian fit profiles. CS
3--2 and \htco\ 2$_{(1,1)}-1_{(1,0)}$ spectra are obtained towards the
position of M1, while the other spectra are obtained towards the
peak of the molecular cloud associated with the SNR. Intensity scale is in main-beam temperature $T_{mb}$.) units.}
\label{fig-mosaic}
\end{figure*}

\begin{figure*} 
\centering
\includegraphics[height=20cm]{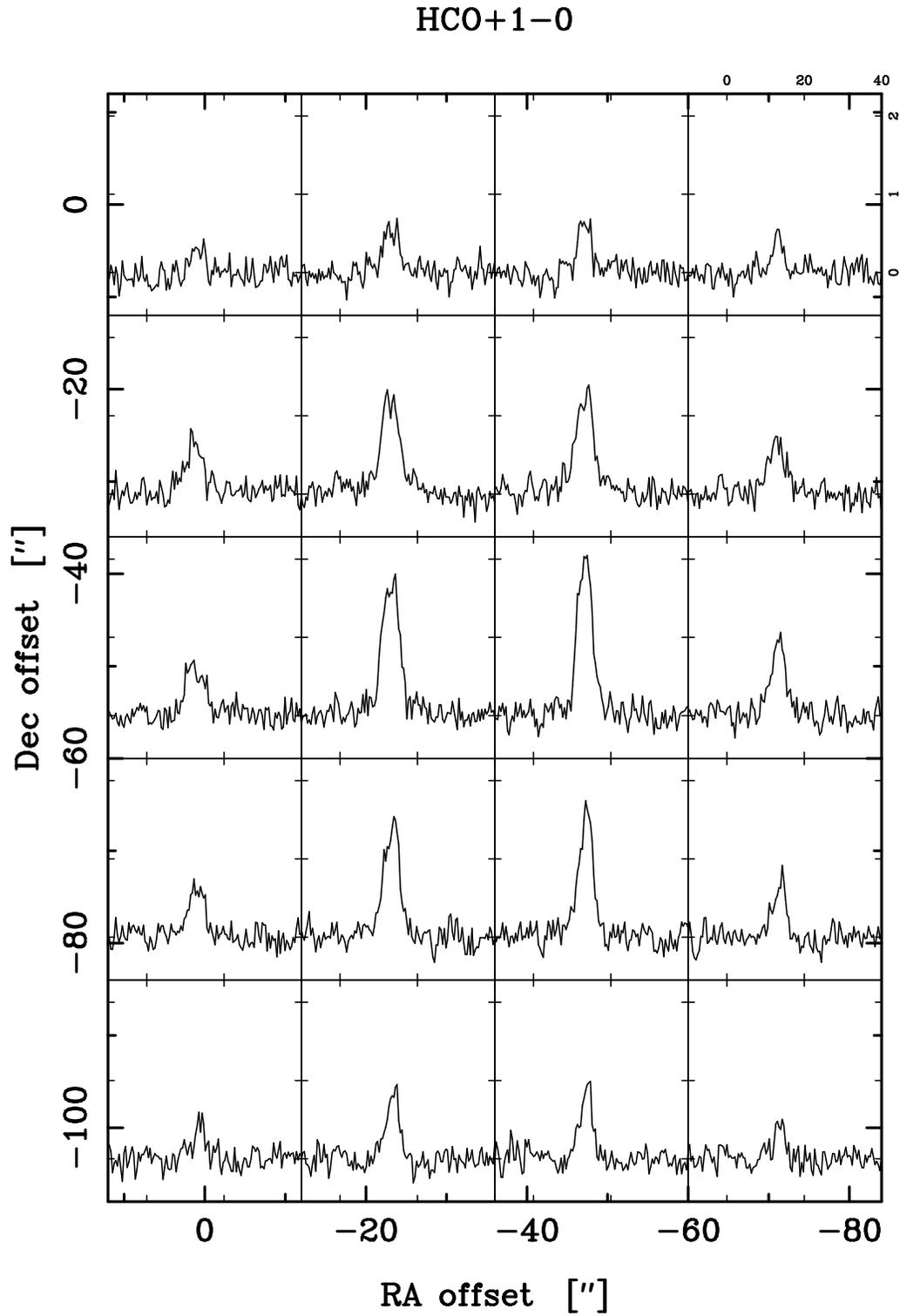}
\caption{The observed HCO+ 1--0 line profiles of the +16\kms\
molecular cloud towards \candy. The line profiles show asymmetry and flattened profiles, suggestive of self-absorption. Intensity scale is in main-beam temperature $T_{mb}$. }
\label{fig-hco+}
\end{figure*}

Several molecular clouds were detected along the line 
of sight to \candy. Figure~\ref{fig-features} shows a sample 
 \co\ 1--0 spectrum towards the SNR  in the velocity range $-40$ to +80\kms.  There 
 are features seen at $-10$, +6 and +16\kms. 
To identify the molecular cloud associated with the SNR, we have
assumed that the cloud has a velocity close to  the maser velocity, i.e., $+16$\kms.
Figure~\ref{fig-mosaic} shows the molecular-line spectra for a variety of transitions and
species in the velocity range $-10$ to $+40$\kms, all measured towards the
 peak position of the molecular cloud described above, except for the CS 3--2 and \htco\
 $2_{(1,1)}-1_{(1,0)}$ transitions, which were measured towards the maser position M1. 
Since the \tco\ spectra taken with the Mopra telescope have different grid positions 
and intervals to those from the SEST data, the \tco\ 1--0 spectrum  shown in 
Figure~\ref{fig-mosaic} is from the pointing at ${\rm RA}(2000) = 17^{\rm
 h}~17^{\rm m}~58\fs5$, ${\rm Dec.}~(2000)=-37\degr~26\arcmin~10\farcs
1$, which is the closest grid point to the SEST peak position. Most of the
spectra appear to  have symmetric, Gaussian-like profiles.
The line profile of HCN 1--0 shows all three hyperfine
components and was fitted with three Gaussian profiles. However, some spectra do 
show asymmetric profiles and flattening, indicative of self-absorption in the molecular 
cloud. The HCO+ spectra, shown in Figure~\ref{fig-hco+}, demonstrate this well. 
The self-absorption is most likely caused by cold gas in front of the warmer gas 
associated with \candy, as seen in other SNRs \citep[eg][]{vand93}.
The basic line parameters derived from the fit to 
the spectral profiles  are summarised in
Table~\ref{tab-candy-sest}. The spectra have centre  velocities
between 12.9 and 16.7\kms, which is in good agreement with the OH
 maser velocities in Table~\ref{tab-masers}. The line widths range
from 2.7 to  6.5\kms, with an average value of $\approx 4$\kms. 

\begin{figure*} 
\centering
\includegraphics[height=10cm]{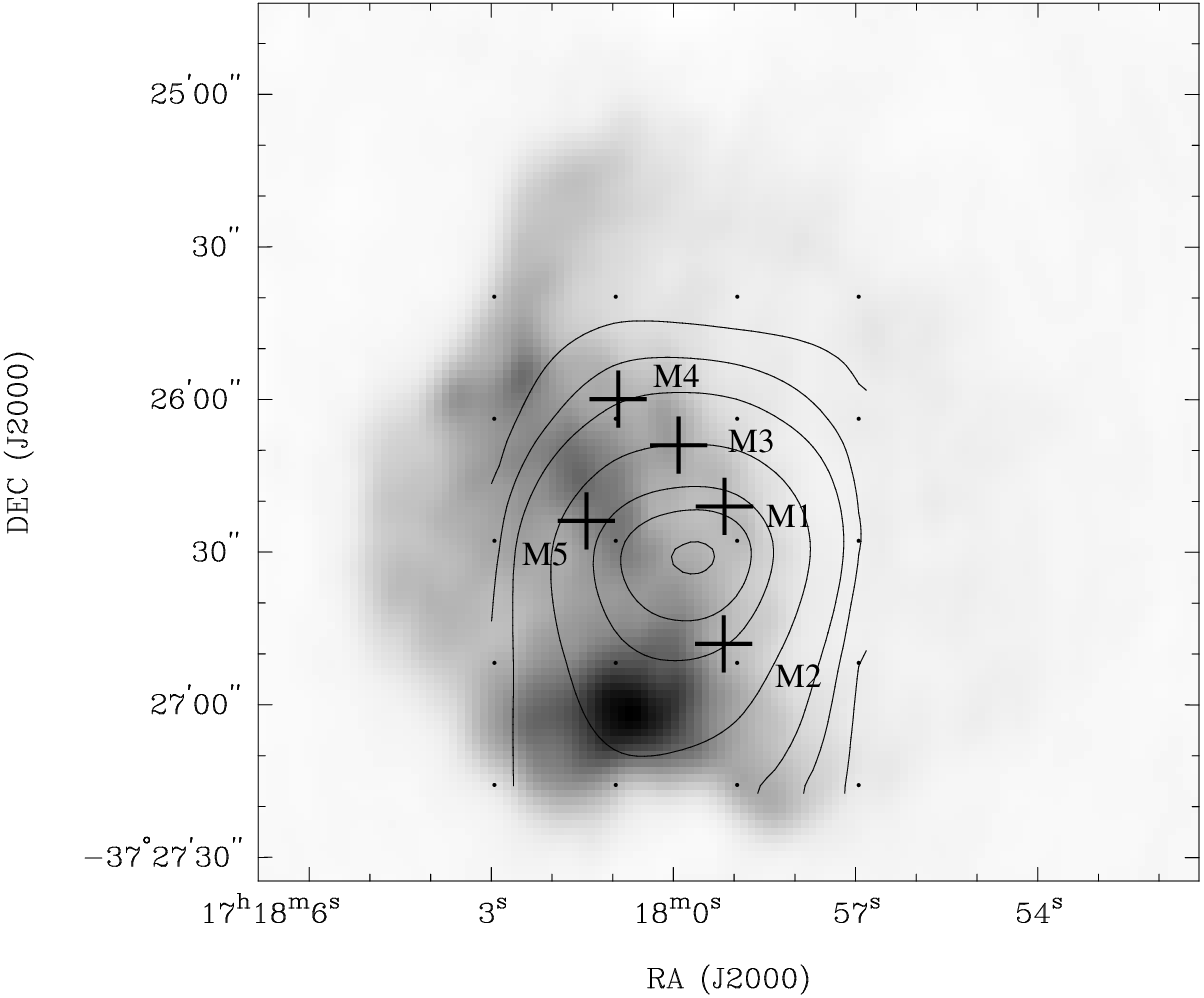}
\caption{Contours of the velocity-integrated (10--20\kms) \co\ 1--0
emission obtained with the SEST with 45~arcsec resolution, overlaid on the 18 cm greyscale radio
continuum obtained with the ATCA. The contour levels are:
16, 24, 32, 48, 63, 71, 78 K\kms. The dots mark the grid positions of the SEST
observations. The crosses  mark the \oh maser positions.}
\label{fig-20+co}
\end{figure*}

Figure~\ref{fig-20+co} shows contours of velocity-integrated
(10--20\kms) \co\ 1--0 emission  obtained with the SEST overlaid on
the 18~cm continuum image of the SNR. The molecular gas peaks at
position ${\rm RA}(2000)=17^{\rm h}~17^{\rm m}$~59\fs6,  
${\rm Dec.}~(2000)=-37\degr~26\arcmin~32\farcs 0$, which is close to the position 
of maser M1. Molecular maps of \co\ 2--1 emission, as well as \tco,
\hcop\ and HCN emission (not shown), are remarkably similar to the \co\
1--0 image in Figure~\ref{fig-20+co}. Our results are consistent with the \co\ maps 
of  \citet{dubner04}, also obtained with SEST. For the CS, \htco\ and SO
observations, emission was strongest towards maser M1. No emission
was detected from the molecule SiO.

%%%%%%%%%%%%%%%%%%%%%%%%%%%%%%%%%%%%%%%%%%%%%%%%%%%%%

\subsection{Properties of the molecular gas} 
\label{sec-mol-gas}

To estimate the physical properties of the molecular gas we used  
 a statistical-equilibrium excitation code RADEX\footnote{Now also 
available on-line at http://www.strw.leidenuniv.nl/$\sim$moldata/radex.html} supplied 
 by J. H. Black (private communication) to model the molecular
line emission. The code uses a  mean escape probability (MEP) 
approximation for radiative transfer \citep*[for more details see][]{jan94,vand07}.  
For a given kinetic temperature and density of the gas, together with
a total molecular column density and species
line width, the code calculates the molecular line intensities, which 
can be compared with observed values. 

\co\ lines are commonly used to derive the kinetic 
temperature of the gas because they are easily thermalised (\tex =\tkin), even at low 
densities. The estimated \co\ 1--0/2--1 line ratio differs from unity, which implies that self-absorption 
might be present and/or  that there are molecular clumps with different filling factors for the 
two sets of observations  (the  beam sizes range from 20~arcsec to 55~arcsec). Indeed, we 
have already established that our observations show some degree of  self-absorption. In 
addition, \h\ emission is typically detected in highly clumped gas and our higher frequency 
data will also be subject to beam dilution effects. To be able to 
compare our line intensities for the same molecular species, a correction must be made for 
the assumed or derived source size. We adopt a source size of 
$40\times90$ arcsec$^2$ for the \co\ cloud, which is the the size estimated from both the \h\ 
and OH emission (see below). This coincidence implies that the value could be the true 
size of the cloud interacting with the SNR.  The actual source brightness temperature,  
corrected for the source size, can be found from
 $T_S=T_{mb}(1+\theta{^2}{_B}/\theta{^2}{_S})$, where $\theta_B$ and $\theta_S$ are
the beam and source FWHM, respectively, and $T_{mb}$ is the observed main-beam 
brightness  temperature \citep[e.g.,][]{rohlf96}. We derive
 $T_S($\co$~1-0) \approx 28$~K and $T_S($\co$~2-1) \approx 36$~K, and 
a  \co\ 2--1/1--0 line intensity ratio of $\approx 1.3$. 
 Importing these values into the RADEX code and adopting an uncertainty 
of $10\%$ in $T_S$, we derive a kinetic 
temperature of $T_{kin}\approx 60\pm15$ for a gas density 
 of \e{4}\cm{-3}, or $T_{kin}\approx 45\pm5$ for a gas density 
 of \e{5}--\e{6}\cm{-3}. These values may be lower limits because of 
the presence of self-absorption. The measured line ratio 
 of \co\ to \tco\ 1--0 main-beam brightness temperature was 
$\approx 4.5$, which implies that the \co\ emission is optically thick. In this case,
the derived kinetic temperature refers to the region from which the
optically thick \co\ emission arises, which might not apply to the
 entire molecular cloud. Using an overall $^{12}{\rm C}$ to
$^{13}{\rm C}$ ratio of $50\pm20$ \citep[e.g.,][]{langer90}, 
we estimate an upper limit for the \co\ 1--0 optical depth of $\approx 16$ and for 
\tco\ 1--0 an  optical depth of $\approx 0.3$.

To derive the density of the molecular cloud it is common to use
molecular species with a dipole moment larger than that of \co, such as
CS. The distribution of  the CS 2--1 emission 
(not shown here) implies that CS emission is concentrated in a much
smaller region than for the \co\ emission, so  a  source size of $\approx 35$~arcsec 
is used to correct the CS brightness temperatures for beam dilution effects. Modelling 
the two CS transitions with kinetic temperatures of $\approx 45$\,K implies a gas density 
of  $n({\rm H}_2) \approx 1.6\times 10^5$\cm{-3}. 
We note that our analysis is sensitive to the line ratio of the 
transitions used, which in turn depends on the adopted source size. 

The excitation analysis implies a \co\ column density 
of $\approx 4$\ee{17}\cm{-2}. Applying the isotope abundance ratio of $\approx 50$,
given above, we obtain  a \tco\ column density of
$\approx 8\times10^{15}$\cm{-2}. The CS column density
derived from the excitation analysis is $\approx 3\times10^{13}$\cm{-2}.
 The non-detection of the CS 5--4 line is also consistent with these physical parameters for
 the gas. We assume that other molecular transitions have a small opacity. In fact, the ratios 
 of the hyperfine lines  
 in HCN, (F=0--1)/(F=2--1)$\approx$1.3 and  (F=1--1)/(F=2--1)$\approx$2.9, deviate 
 significantly from the LTE values of 0.2 and 0.6 \citep[eg,][]{chin97}, respectively, suggesting 
 that HCN is optically thin. For an estimation of the total column density for the molecular 
 cloud associated with the SNR we use the standard fractional abundance
of \co\ relative to \h\ of \e{-4}\  \citep[e.g.,][]{irvine87,vand93}, which implies 
$N$(H$_2) \approx 4\times 10^{21}$\cm{-2}.
For  the other molecules species observed  we used the relation for 
optically thin transitions \citep[e.g.,][]{rohlf96}: 
\begin{equation}
N_l = 2.07\times 10^{3} \frac{g_l\,\nu^3}{g_u\,A_{ul}} \int{T_S dv},
\label{eq-n_mol}
\end{equation}
 where $g_i$ ($i=u,l$) is the number of states with 
the same energy ($g_i=1+2i$), $A_{ul}$ is the Einstein 
$A$-coefficient of spontaneous emission, and $\nu$ is transition
frequency in units of GHz. From this relation we derive the 
column densities for \hcop\ and HCN, using the same source size as for
\co, and for SO using the source size for CS.  The values  
 are listed in Table~\ref{tab-abundances}.

\subsection{\h\ emission}

\begin{figure*} 
\centering
\includegraphics[height=7cm]{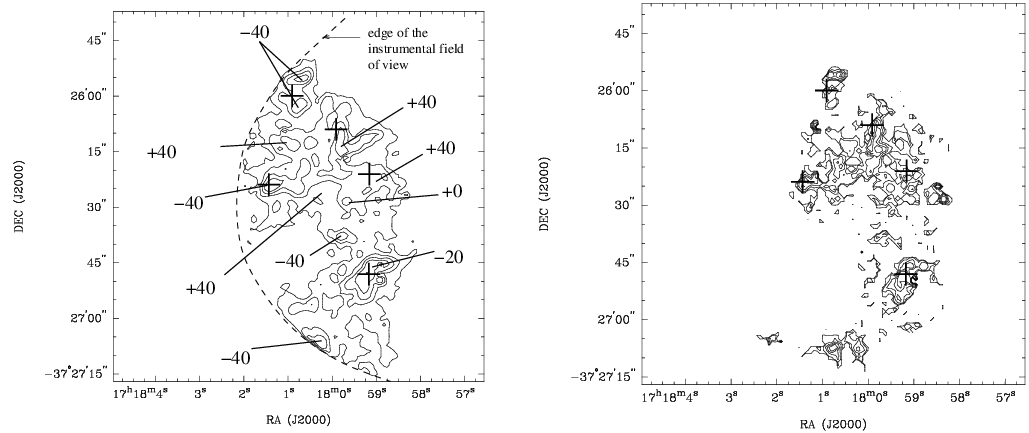}
\caption{Contours of the velocity-integrated \hone line emission
({\it left}) and \htwo line emission ({\it right}). Note that flux
densities are not corrected for extinction (which is assumed to be
 a minimum 3\,mag in the $K$ (2.12$\micron$) band). The contours are for
1--0~S(1) emission: 3.3, 6.7, 13.2, 19.7 and 26.3
\ee{-5}\ergs; for 2--1~S(1) emission: 2.7, 3.6, 4.5, 5.3 and 6.2
\ee{-5}\ergs. The 1$\sigma$ noise in the 1--0
line is 6\ee{-6}\ergs\ and in the 2--1 line is 4\ee{-6}\ergs. 
The crosses  mark the \oh maser positions. Line-centre
velocities of \h\ emission are also indicated in the left panel.}
\label{fig-candy-h2}
\end{figure*}

\begin{figure*} 
\centering
\includegraphics[height=10cm]{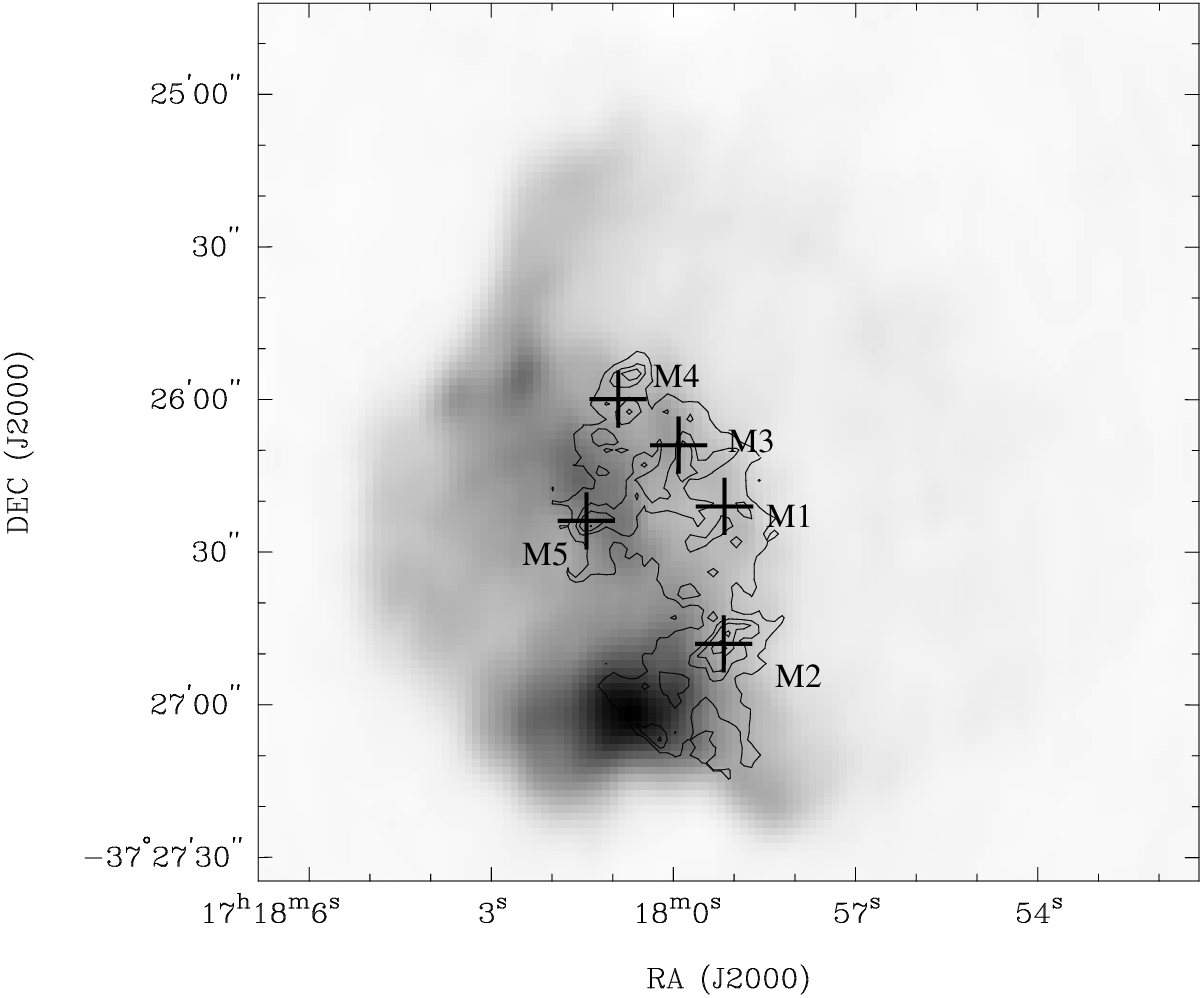}
\caption{Contours of \hone emission overlaid on the 18 cm greyscale
radio continuum image of G349.7+0.2. The contour levels (not corrected
for extinction) are: 3.2, 9.6, 16 and 26 \ee{-5}\ergs. The crosses mark the \oh maser positions, which are again shown numbered.}
\label{fig-h2+20}
\end{figure*}

Figure~\ref{fig-candy-h2} shows the velocity-integrated 1--0 and 2--1
S(1) \h\  line emission detected  towards \candy. The 1$\sigma$ noise in the 1--0
line is 6\ee{-6}\ergs\ and in the 2--1 line is 4\ee{-6}\ergs. The
south-eastern part of the \hone  emission is truncated by the instrumental
field of view, but the total extent of  the emission can be determined from
the \htwo image.  The \h\ emission extends about 1.5 arcmin
with a maximum width of 40 arcsec. It has a clumpy  structure
containing several peaks (knots), which are about 15 arcsec in
size. In general, the maser positions are located near 
these \h\ knots.  The line fluxes of the knots are corrected 
for a minimum extinction of $A_K\approx 3$~mag
because the source is located 
beyond the Galactic Centre. The values are summarised in
Table~\ref{tab-h2} and the numbering of the knots is the same
as for the masers. After correction for extinction, the
intensity ratio of the 1--0 and 2--1 S(1) \h\ transitions is calculated to be 
about 5--6. However, we note that because of the uncertainties in the absolute
calibration at the two line frequencies, the derived line ratios may
have a (constant) scaling error of up to 50\%. There is a range
of line centre velocities across the source, as indicated in 
Figure~\ref{fig-candy-h2}. Most of the \h~emission has a line centre velocity 
 around $-40$\kms. However, the emission from \h\ knots 1 and 3
 have a peak line center velocity of $+40$\kms,
while the  emission from knot 2 has a peak velocity of $-20$\kms. As mentioned 
previously, the velocity resolution is low ($\approx 75$\kms) and we cannot determine 
the peak velocities very accurately ($\approx 20$\kms uncertainty). Nevertheless, we 
can report that \h\ emission was detected in the velocity range of $-$40 to +40\kms,  
which encompasses both the velocites of the \oh masers and the parent molecular cloud,
at roughly 16\kms. The velocity-integrated \h\ emission contours are superimposed  
on the 18 cm radio continuum greyscale image  in Figure~\ref{fig-h2+20}. We note that 
there is also a clump of \h\ emission coincident with the radio continuum peak, which is
clear in the 2--1~S(1) line image, but has been truncated by the edge of the field of
view in the 1--0~S(1) line image.

\subsection{Search for ground- and excited-state OH maser emission} 

\begin{figure*} 
\centering
\includegraphics[height=6cm]{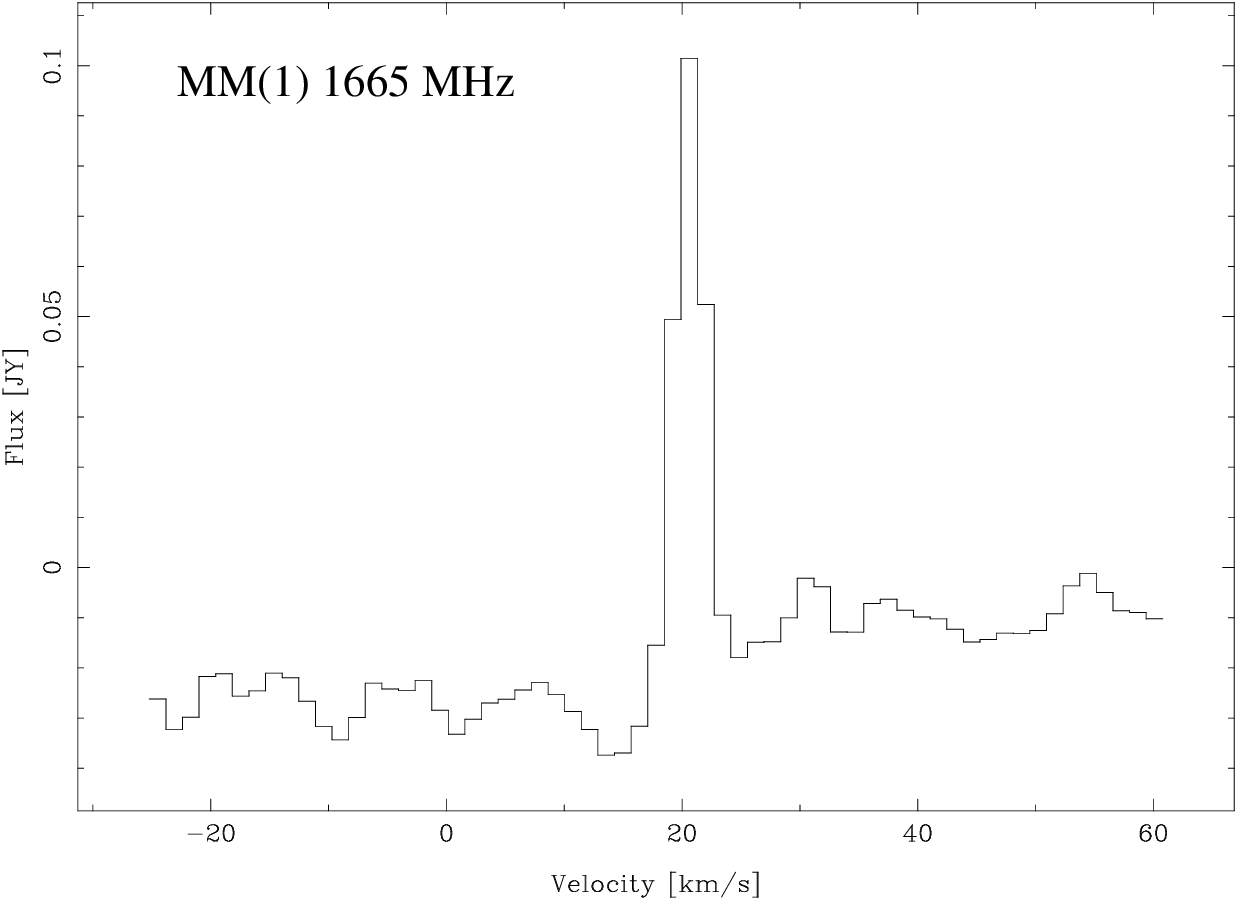}\\
\includegraphics[height=6cm]{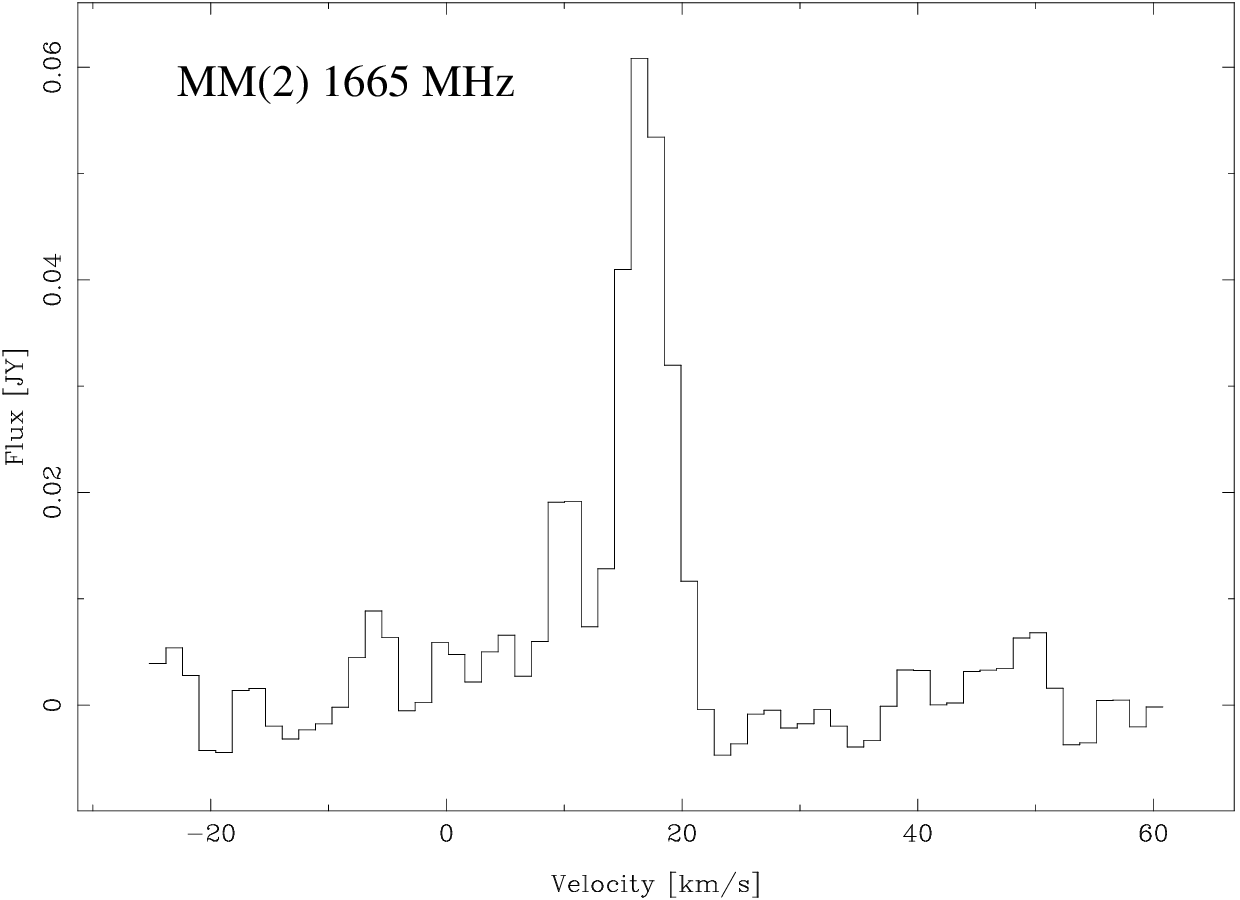}\\
\includegraphics[height=6cm]{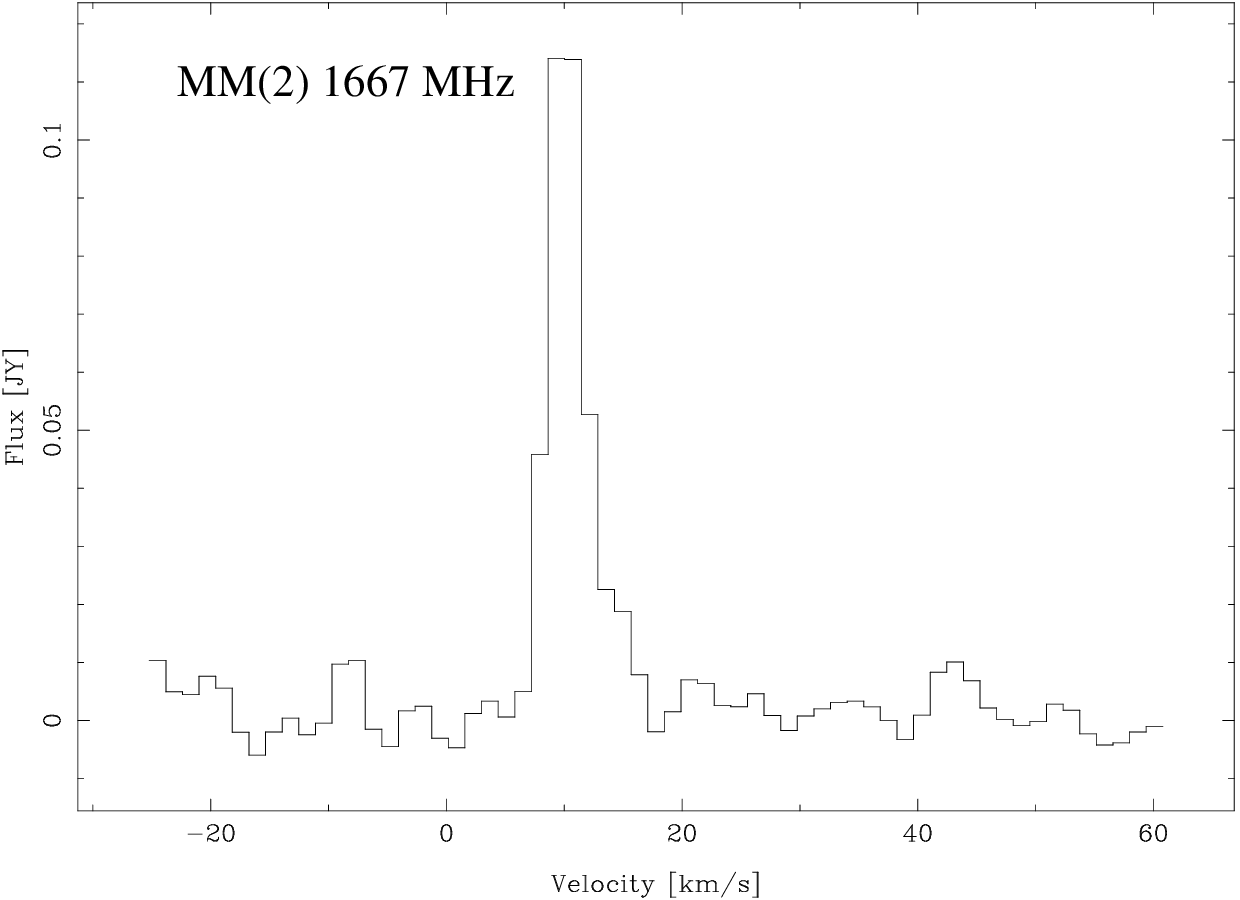}\\
\caption{Main-line maser profiles towards two locations in
\candy. MM(1) is located within the SNR contours, and MM(2) originates
from the UC \hr\ region located southeast from the remnant.}
\label{fig-oh-mm}
\end{figure*}

We have detected weak emission from the OH main-line
masers at 1665 and 1667~MHz towards two locations in the observed field. 
The positions and  flux densities of the sources are summarised in
Table~\ref{tab-mainl} and the maser profiles 
are shown in Figure~\ref{fig-oh-mm}. One  of the main-line masers,
denoted MM(1), is located in the direction of the SNR. Its position is closest (offset 
by 5~arcsec in RA and 1~arcsec in Dec) to that of  maser
 M3 and emission was detected only in the  1665~MHz transition. The 
second  maser, MM(2), was detected in both main-line transitions 
and is centred on the UC H\,{\sc ii} region IRAS~17147--3725.
 Therefore, MM(2) is clearly related to star-formation. The peak velocity 
of the 1665~MHz transition of 16\kms\ is somewhat different  from the peak velocity 
of the 1667~MHz transition of 10\kms, but both velocities fall within the range for 
the molecular gas found towards the H\,{\sc ii} region (10--20\kms). This discrepancy in
the peak velocities of the masers may occur in star-forming regions
\citep[e.g.,][]{masheder94,caswell99}, indicating a slight difference in their formation sites 
in the molecular cloud and consequent variations in the conditions producing the masers. 

To date, main-maser lines have not been found detected coincident with  \oh\ masers in 
SNRs. There is an unresolved {\em IRAS} source IRAS 17146--3723 located
$\approx 15$~arcsec away from MM(1), but the angular resolution
 for the {\em IRAS} observations ranges from 0.5--2~arcmin 
between 12 and 100\micron. Both \citet{estela2001} and \citet{dubner00} considered whether this source is 
dust heated by the SNR shock or a protostellar object. 
To investigate the nature of MM(1) towards \candy, we examined the available IR 
data from the Two Micron All Sky Survey (2MASS) Point Source Catalog \citep{skr2006} 
and the {\em Spitzer Space Telescope} GLIMPSE Catalog \citep{ram08}. There is  a 
GLIMPSE source  SSTGLMC~G349.7294+00.1747 detected at 5.8\micron\ (8~mag) and 
8.0\micron\ (6.5~mag), 
located 0.8 arcsec from MM(1), but with no 2MASS counterpart. For comparison, there is a 
GLIMPSE source SSTGLMA~G349.7213+00.1208 located 0.7 arcsec from MM(2) 
 with 6.5~mag at 5.8\micron\ and 4.8~mag at 8.0\micron\ , which does have a 2MASS 
 counterpart. Thus, it would appear that MM(1) may also be related to star formation. 
 However, detection of a possible H{\sc ii} region associated with MM(1) and the GLIMPSE 
 source would be very difficult because the radio continuum emission from the SNR is very 
 bright. As evidence, the nearby UC H{\sc ii} region  IRAS~17147--3725 has a brightness 
 comparable  with the weakest emission measured from the SNR. 

Our search for the excited-state maser lines was unsuccessful and the limits given by
the 1$\sigma$ rms values are listed in Table~\ref{tab-atca}. 
The relationship between ground- and excited-state transitions of
OH masers  has been studied in detail in star-forming regions
 observationally and theoretically \citep*[see e.g.,][ and 
references therein]{gray92,macleod97,pavlakis96b,pavlakis00}, and recently 
in molecular gas associated with SNRs \citep{wardle07,pihlstrom08,mcdonnell08}. 
\citet{gray92} proposed a correlation between
 masing at 1720 and 4765~MHz in warm gas (\tkin\ $> 100$~K), and
between 1720 and 6035~MHz in cooler molecular gas, which
was confirmed using single dish observations of star-forming regions \citep{macleod97}. 
 Models by \citet{pavlakis96b,pavlakis00}
imply that OH transitions around 4.7~GHz are only excited in gas
with higher density ($> 10^{6}$\cm{-3}) than is required to produce
the 1720 MHz maser. Furthermore, the transitions at 
6031 and 6035~MHz require a
strong far-infrared (FIR) radiation field. A 
satellite-line maser at 6049~MHz is
more likely to be found, as it requires physical conditions not dissimilar to those 
needed for production of  the 1720~MHz line, but it is expected to be very weak 
at densities $< 10^{6}$\cm{-3} \citep{pavlakis00}, and would 
probably lie below our detection limits. Our results are
 therefore consistent with the OH excitation models.

\subsection{Search for CH$_3$OH maser emission} 

Methanol (CH$_3$OH) masers are related to star-formation and are 
often found to co-exist with OH masers \citep[e.g.,][]{plambeck90,phillips98}. 
We searched for CH$_3$OH maser emission at 6.6~GHz towards 
the location of the two OH main-line masers in the field of \candy. 
 Of particular interest was whether the MM(1) maser had a
 CH$_3$OH maser counterpart, in order to clarify its origin. 
However, no maser emission was detected; the 1$\sigma$ rms limits are listed in
Table~\ref{tab-atca}. The conclusion is that either CH$_3$OH is not sufficiently abundant 
or that the local conditions (gas temperature, density, dust
temperature) are such that only OH masers are favoured \citep*{cragg02}.

\subsection{OH absorption}

\begin{figure*} 
\centering
\includegraphics[height=7cm]{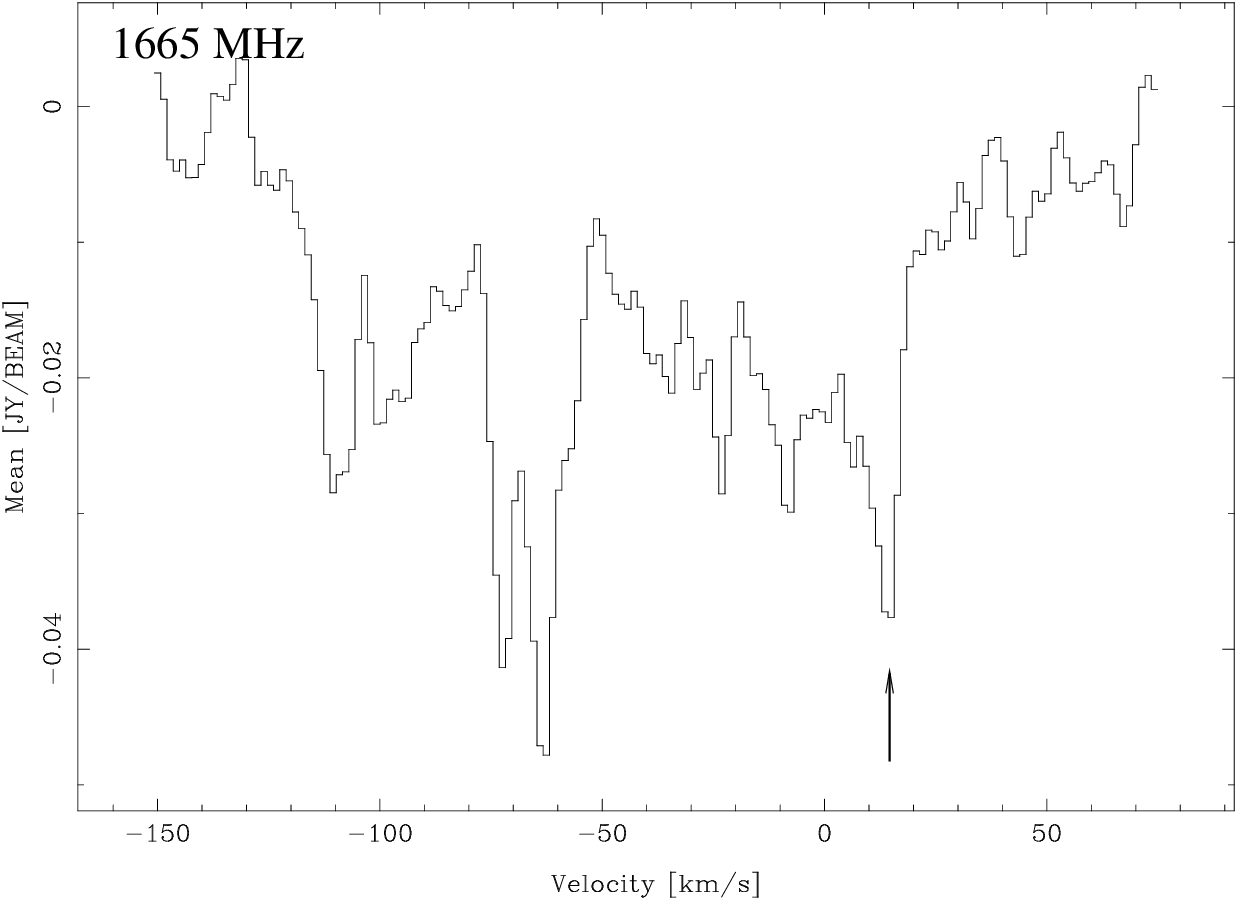}
\includegraphics[height=7cm]{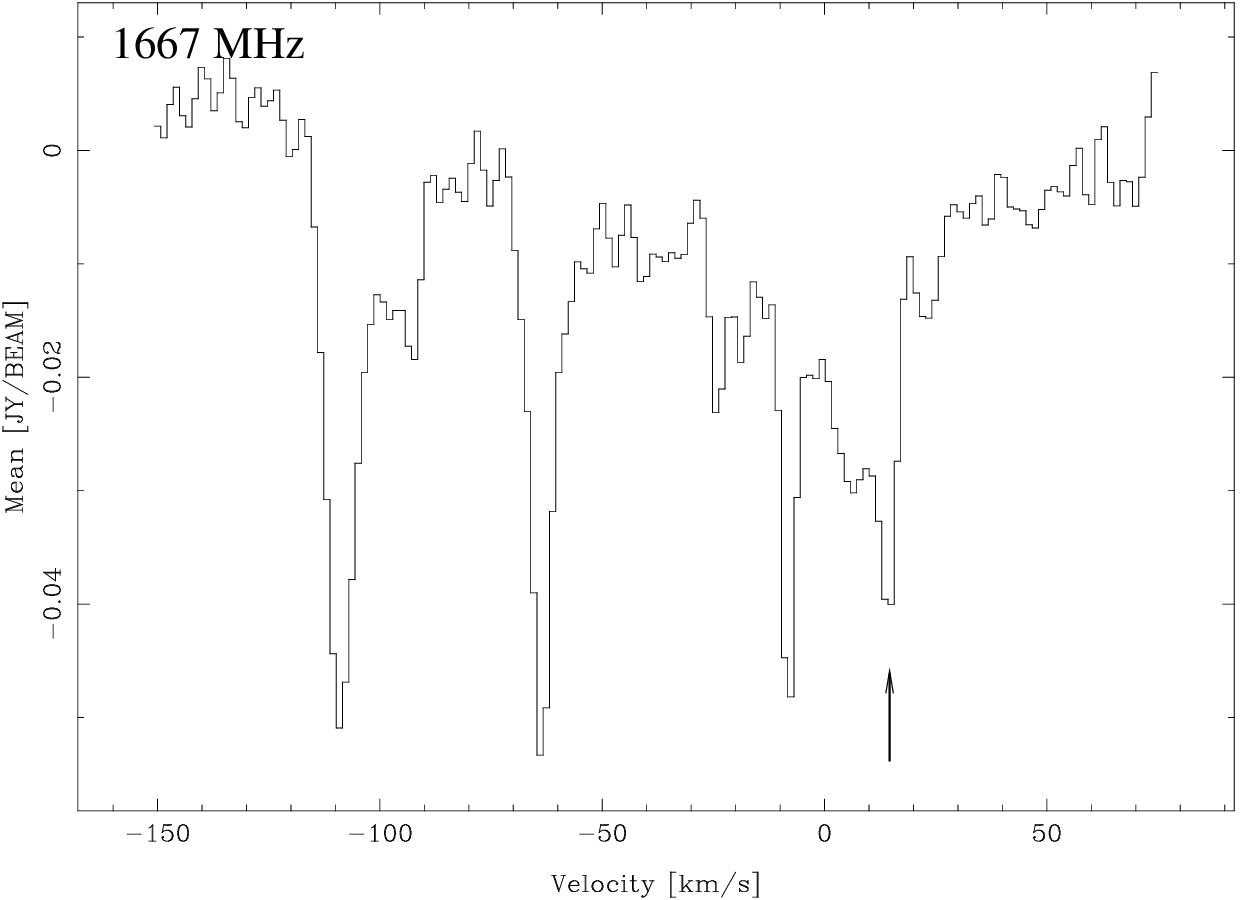}
\caption{Line profiles of 1665 and 1667~MHz OH absorption, Hanning
smoothed by 3 channels, show the features along the line of 
sight to the SNR \candy. The arrow indicates the velocity of the
material associated with the remnant.}
\label{fig-oh-abs}
\end{figure*}

\begin{figure*} 
\centering
\includegraphics[height=8cm]{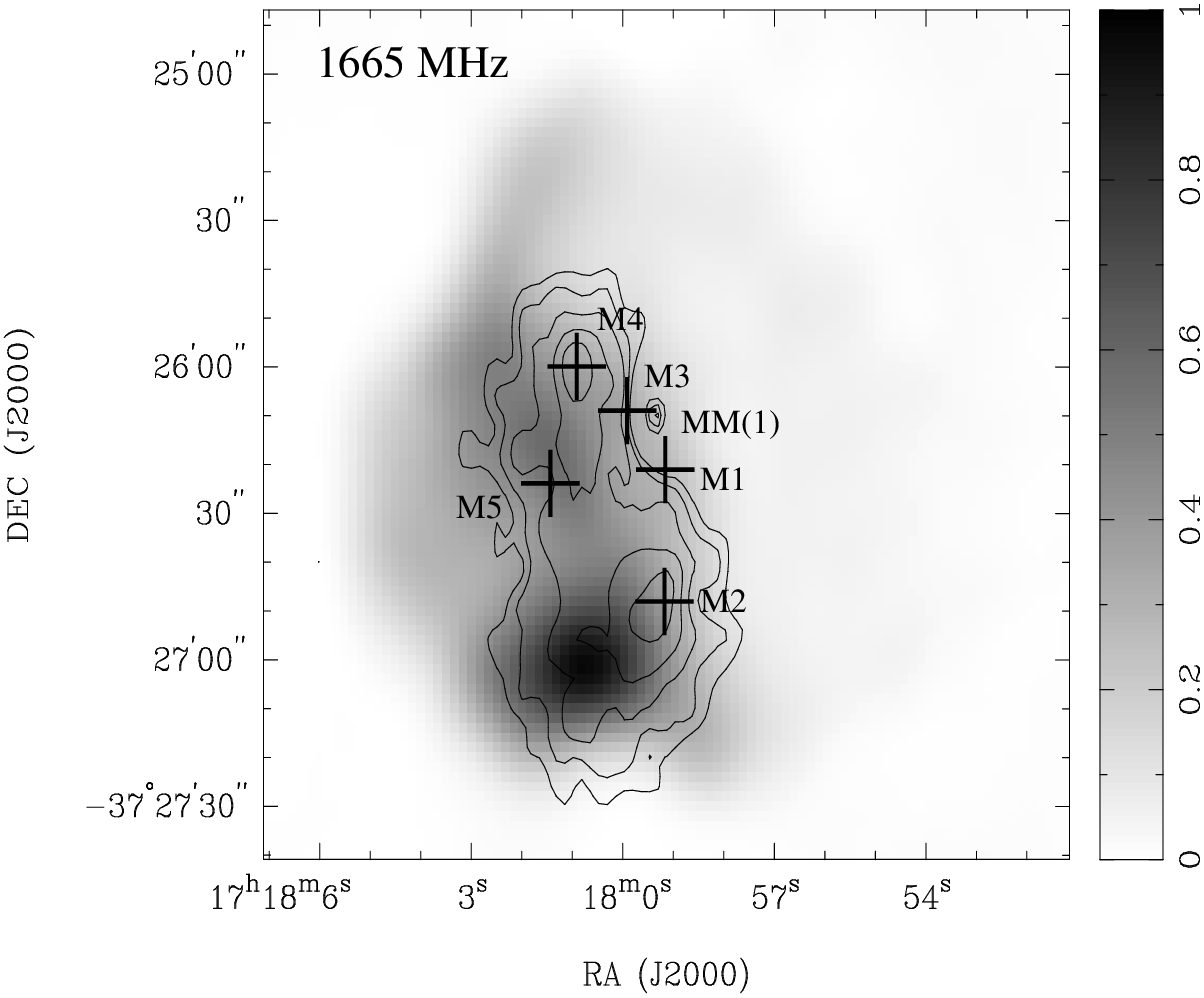}\\
\includegraphics[height=8cm]{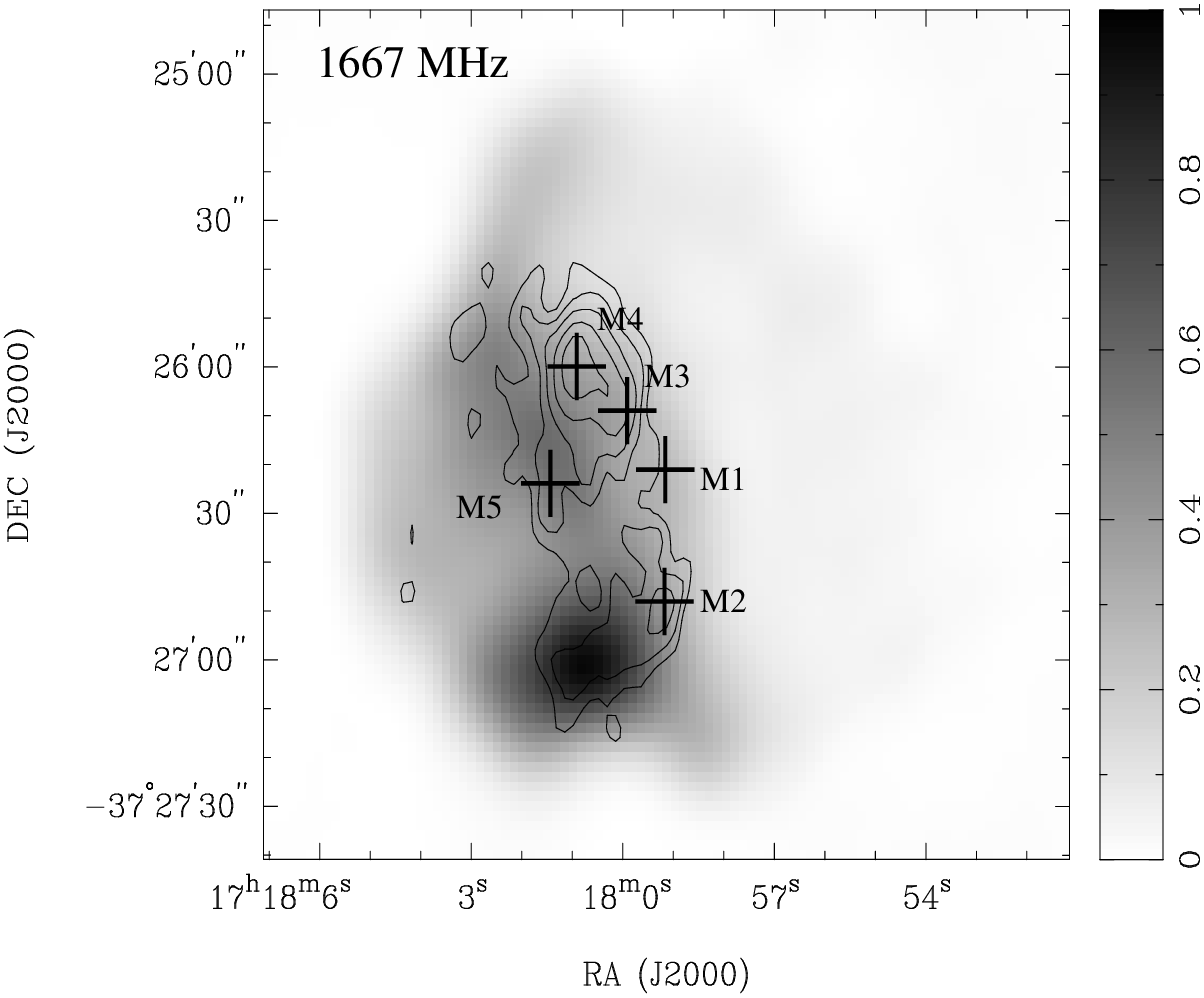}
\caption{Contour images of the 1665 and 1667~MHz OH absorption, integrated 
between 10 and 20\kms\ overlaid on the greyscale 18 cm 
continuum image of G349.7+0.2. The contours are: $-0.13, -0.16, -0.20,
-0.25$ and $-0.30$~Jy\,beam$^{-1}$\kms. The crosses  mark the \oh
maser positions and the compact source in the 1665~MHz image is an
 OH(1665~MHz) maser, marked as MM(1).} 
\label{fig-oh+20}
\end{figure*}

The 1665 and 1667~MHz OH absorption profiles toward \candy\ are shown in
Figure~\ref{fig-oh-abs} in the velocity range $-150$ to $+70$\kms, Hanning smoothed 
over 3 channels. Prominent features are present around $-110, -95,
-65, -25, -10$, $+6$ and +16\kms.  The 1665~MHz profile
differs from the 1667~MHz spectrum in having an additional feature at
$-70$\kms. This feature originates from a bright maser located
north-west of the SNR, outside the observed field of view. These OH absorption features are also  
seen in H\,{\sc i} absorption \citep{casw75}, and some of them (for which we
have velocity coverage) are seen in our SEST \co\ spectra, i.e., features at $-10$, $+6$ 
and +16\kms.  The +16\kms\ feature, which we associate
with the SNR because of the common velocity with the \oh masers, 
is partially blended with the +6\kms\ feature. 
 To investigate the distribution of the 16\kms\ OH cloud  we produced 
 velocity-integrated (10--20\kms) images of OH absorption with a resolution of 
15\arcsec$\times$12\arcsec\ (P.A. = $-$3\fdg2), degraded from the best available 
resolution to improve the image sensitivity. The OH images are shown in 
Figure~\ref{fig-oh+20}, overlaid on the 18~cm radio  continuum image. The distribution 
of the OH cloud does not  mirror the continuum emission, which would occur for a uniform 
overlaying absorption cloud, but appears as an
elongated feature,   40\arcmin\ wide and 1\farcm5 long, 
covering the region of the \oh maser emission. We note that for the section of the SNR  
containing the masers the continuum emission is reasonably constant, which suggests 
that the variation in OH absorption is indicative of variations in OH column density. 
 The two troughs in the OH absorption distribution coincide with the M2 and M4 maser 
 locations. The 1665~MHz distribution is more extended than that of 1667~MHz, 
 and there is an indentation in the OH absorption distribution near the compact 
 MM(1) maser, shown in Figure~\ref{fig-oh+20}.

\begin{figure*} 
\centering
\includegraphics[height=5cm]{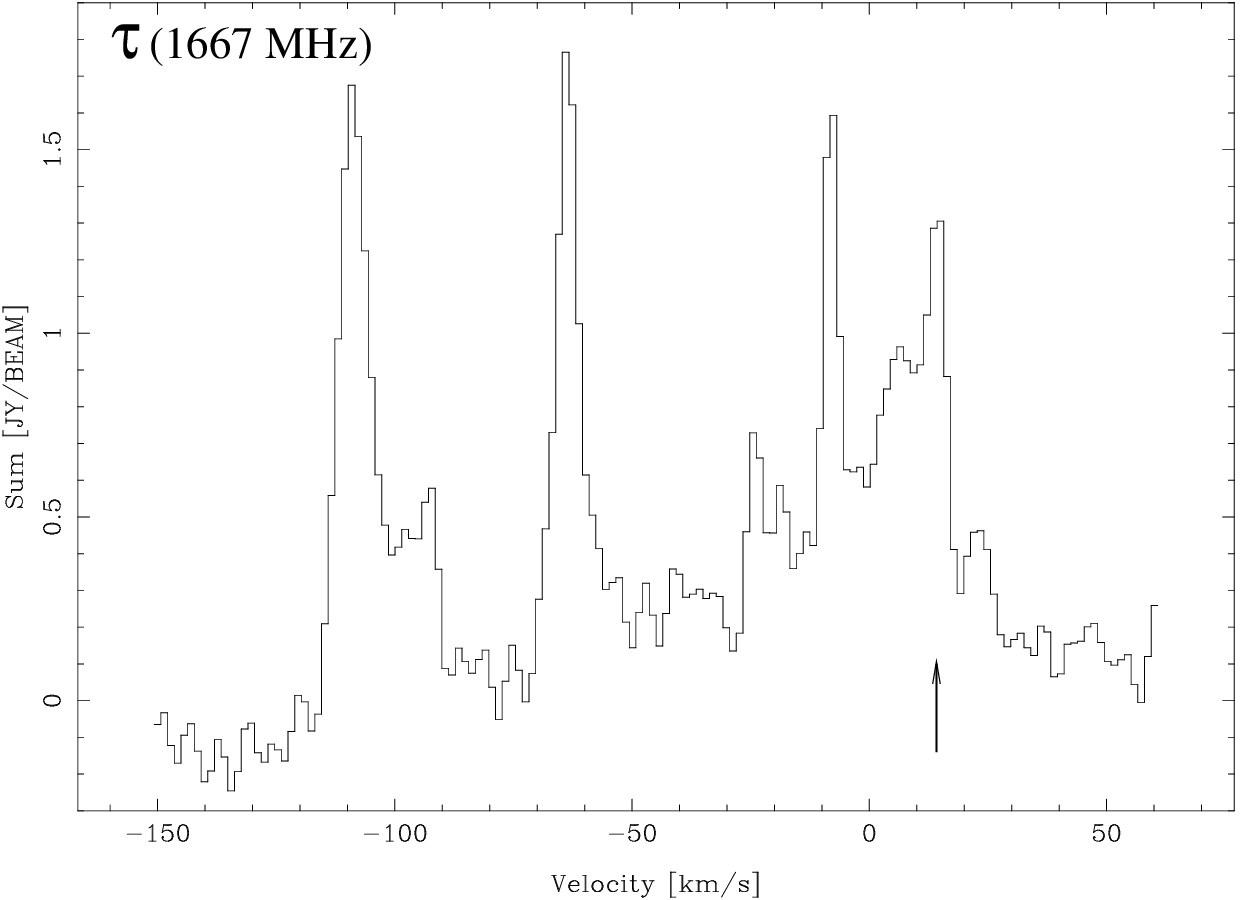}
\includegraphics[height=5cm]{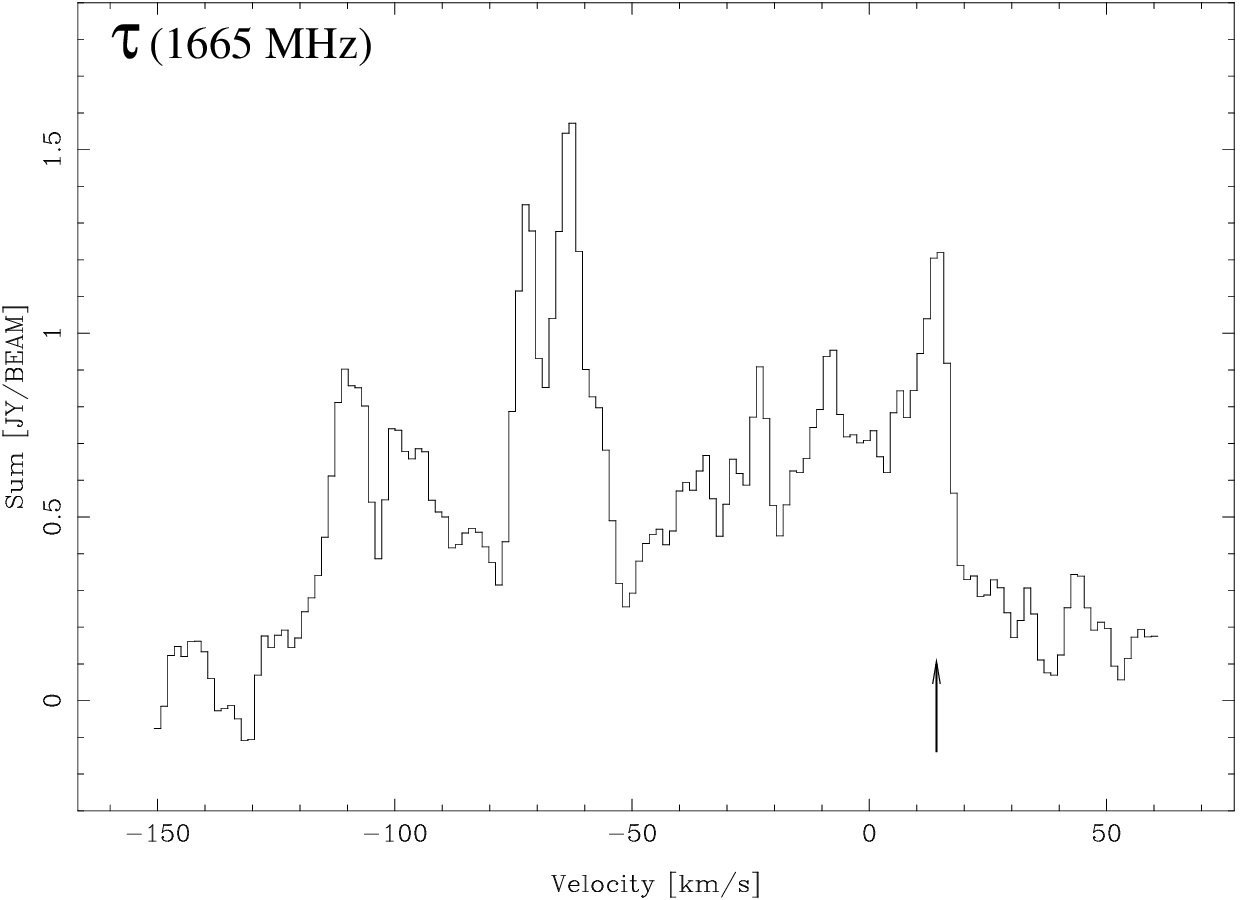}
\includegraphics[height=5cm]{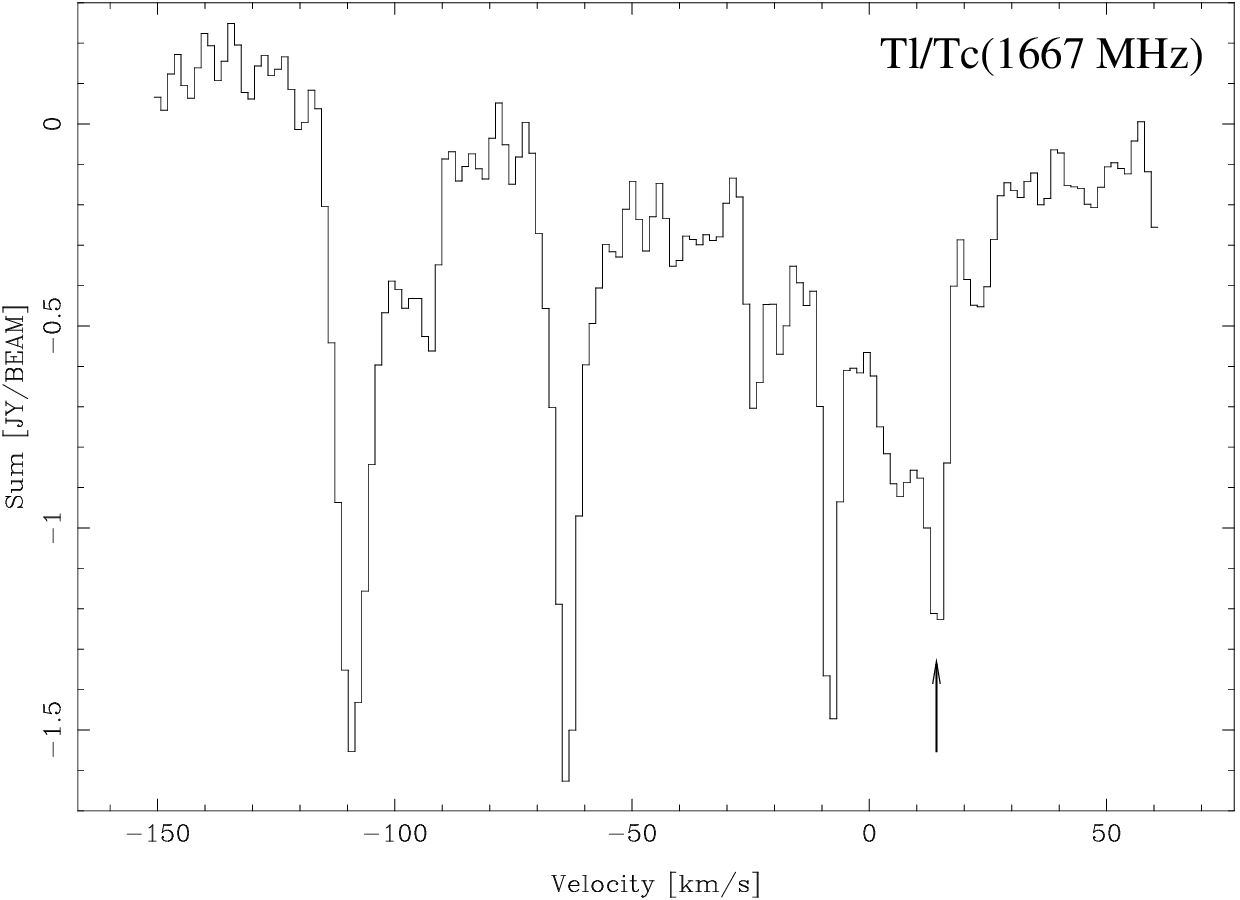}
\includegraphics[height=5cm]{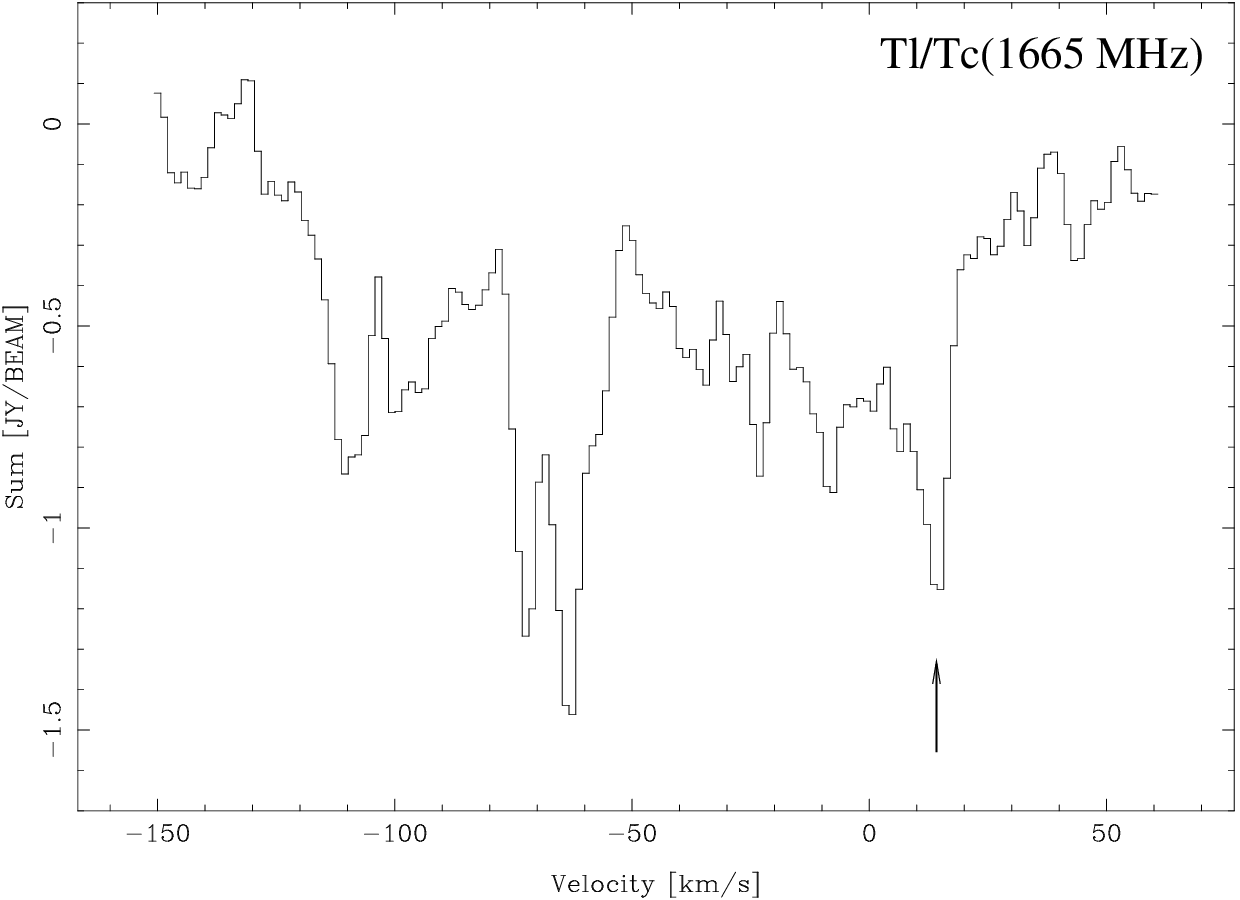}
\caption{Line profiles of the 1665 and 1667~MHz OH optical depth (upper
 panels), and profiles of the line-to-continuum ratio (lower panels)
towards the peak of the OH cloud. The spectra
have been Hanning smoothed over 3 channels. The arrow indicates the velocity of the
material associated with the remnant.}
\label{fig-tau}
\end{figure*}

The optical depth $\tau=-{\rm ln}((T_L/T_C)+1)$ was calculated from the spectral line 
channel maps,  where $T_L$ and $T_C$ are the intensities of the spectral 
line and continuum maps, respectively. The calculation were limited to  
regions with a continuum intensity greater than 25\mjb\,  because of the poor signal-to-noise 
figure  at weaker continuum levels. In Figure~\ref{fig-tau} we compare the optical depth 
profiles and  the line-to-continuum ratios  towards the peak of the OH cloud. The optical 
depths are quite high, around 1.2--1.3. The relative intensity of the OH absorption 
in the 1667~MHz and 1665~MHz lines is expected to have a ratio around 9/5 for 
an optically thin gas in local thermal equilibrium (LTE). 
 The line ratio $T_{1667}/T_{1665}$ in \candy\ is $\sim 1.1$, which is significantly different 
 from the LTE ratio ($\approx 1.8$), probably reflecting the high calculated optical depths. 
OH absorption is typically  observed with low optical depths in the Galaxy, but higher values 
 have been found towards some SNRs \citep[e.g.][]{yusef03-w28}.

The OH column density for the LTE case can be obtained 
from \citep[e.g.,][]{crutcher77}:
\begin{equation}
N({\rm 1665~MHz}) = 4.2 \times 10^{14} T_{ex} \int \tau_v dv {~\rm cm}^{-2},
\end{equation} 
\begin{equation}
N({\rm 1667~MHz}) = 2.3 \times 10^{14} T_{ex} \int \tau_v dv {~\rm cm}^{-2}.
\end{equation} 
 Modelling of the OH population levels implies that for \tkin $\ge 20$~K 
 an OH excitation temperature of 10~K is plausible
 \citep{yusef03-w28}.  Using the above values  of the optical depth, 
we derive a peak OH column density of (2.8--4.7)\ee{15}\cm{-2}. These values are 
consistent with the models for  \oh excitation  \citep{elitzur76,lock99}
 and with the model for OH production in C-type shocks \citep{wardle99}.
Modelling  the OH excitation using all four ground-state OH transitions, \citet{hewitt08} found 
that the line profiles from their observations of 15 SNRs with the Green Bank Telescope  are 
consistent with molecular gas with temperatures of 30-100~K, OH column densities of 
\e{15}--\e{17}\cm{-2} and gas densities of $\sim$\e{5}\cm{-3}. \citet{hewitt08} detected 
enhanced 1720~MHz emission in 
\candy, but were not able to resolve any corresponding main-line absorption lines, principally
 because of insufficient sensitivity and the large 7\farcm2 telescope beam at 1.7~GHz.

\subsection{\htco\ absorption}

\htco\ absorption at 4829~MHz is readily found
in H\,{\sc ii}-molecular cloud complexes, where \hr\ regions are
embedded in dense molecular clouds. There have been only a few
such absorption detections towards SNRs 
\citep[e.g.,][]{whiteoak74,slysh80,denoyer83,reynoso02}, where the associated
continuum emission and resultant absorption  is usually much fainter than for surveyed 
H\,{\sc ii} regions.  We did not detect \htco\  absorption
towards \candy, but there was weak absorption towards the nearby UC
\hr\ region, with an optical depth of $\approx 1$ and a column density of
 $\approx 4.3$\ee{15}\cm{-2} (for an adopted excitation temperature of
10~K and  a line width of 5\kms). For a 3$\sigma$ upper limit and a line-width of 
5\kms, we estimate an optical depth $< 0.013$ and an \htco\ column density  
$< 1.7$\ee{13}\cm{-2} in the direction of the SNR.

\section{Discussion}
\label{sec-candy4}

\subsection{Kinematics and chemistry of molecular gas}

The parameters derived from the millimetre-line observations of 
various molecular species for the molecular cloud associated with \candy of 
$n({\rm H}_2)\approx 10^5$\cm{-3} and  $T_{\rm kin} > 45$~K, are
 consistent with values  required for the production of \oh masers by shock
 excitation \citep{lock99}. However, the measured millimetre-line widths of 
 $\approx$4\kms\ , for all the molecular transitions we observed, 
are smaller than those expected from shocked gas ($\ge 10$\kms).  
Although the shock front containing the \oh masers will be 
seen mostly perpendicular to the line-of-sight (hence showing minimal 
line broadening), some broadening is expected due to deflection of 
the shock in clumpy pre-shock gas \citep{turner91}. High velocity wings were not 
detected in the lines from the  low \co\ transitions, but they may be too weak to be seen 
 with the short integration times used here.  \citet{dubner04} reported 
asymmetry in the \co\ line profiles towards the cloud borders, indicating the line broadening 
and kinematic signatures of shocked gas. We find that the peak velocities vary across the 
source for all the species observed, but there was no particular trend. It is probable that the 
shifts are caused by self-absorption in the spectral lines due to the presence of cold, 
pre-shocked gas, as found in other SNRs \citep[eg][]{vand93,reach05}.  Evidence for this
hypothesis is  the flattened line profiles seen in the \co, \tco\ and HCO+ spectra 
(Figure~\ref{fig-hco+}). The presence of self-absorption in the lines, in conjunction with 
the fact that the various  molecular species sample gas of different densities, could also 
explain the variations in peak velocity between the different molecular species, listed in 
Table~\ref{tab-candy-sest}.

In Table~\ref{tab-abundances} we list the various molecular column densities and
abundances determined using  the \h\ column density of 4\ee{21}\cm{-2}, 
 derived in section \S\ref{sec-mol-gas}. This value differ by an order of magnitude 
from that derived by \citet{dubner00}, but no correction was made of their observed 
intensities for beam dilution, even though the \co\ 3-2 map showed a different distribution 
of molecular gas compared with their 1-0 and 2-1 \co\ maps. The higher transition image is 
quite similar to our \h\ map. Because the angular resolution of our observations is relatively
poor compared with the size scale of the SNR, we were not able to observe the
contributions from ambient and post-shock gas separately. Hence, we are
not able to compare post- and pre-shock abundances in the cloud
associated with \candy. Nevertheless, an attempt has been made to compare our observed 
values with those of the IC~443 from \citet{vand93}, where shock chemistry has been 
studied in great detail. We also compared them with abundances in the dark clouds
 L134N and TMC-1 \citep*{ohishi92}, which were also compared with IC~443 
 \citep{vand93}. Abundances in dark clouds provide information about basic
molecular formation in gas-phase chemistry without contamination
 by ultraviolet (UV) photons from embedded sources. We find that the abundances of CS 
 and HCN are comparable for SNRs and dark clouds. In other words, they remain
unchanged for shocked gas, as expected from the models. 

SiO is the only molecule in IC~443 found to have an abundance enhanced
 with respect to dark clouds \citep{vand93}. Observations
and theoretical modelling indicate that in dark clouds Si is heavily
 depleted because it freezes out onto dust grains. Si-species such as SiO become
 observable in warm regions with temperatures higher than 30~K 
\citep*{ziurys89,mackay96}. Shock models predict enhanced SiO
abundances as a bi-product of gas-phase chemistry when Si is released from the
grains by evaporation or disruption by fast dissociative shocks
\citep{neufeld89,mackay96,schilke97}. This hypothesis is supported by 
observations of young bipolar outflows, where jets from a central star
interact with ambient molecular gas \citep[e.g.,][]{garay98}. We
did not detect SiO towards \candy and estimate an upper limit
 to its column density of  $X$(SiO)$<$7.5\ee{-9}. This limit is much lower than 
the value for IC~443, but is still significantly higher than the limits found for dark clouds. 
More sensitive observations are needed to further examine the SiO abundance in \candy.

Another molecule predicted to be enhanced in shocks
is SO. This enhancement depends on the H/H$_2$ ratio, the
C/CO ratio, and on the initial form (atomic or molecular) of sulphur
\citep[see][and references therein]{vand93}. \citet{pdef93} suggested that SO 
can be found at  distances up to \e{18}~cm from the shock front.  
The telescope resolution at the SO frequency observed is such that beam dilution 
is expected to render any SO enhancement undetectable, unless the species  was 
distributed over a substantial distance. The derived SO abundance of $\approx3$\ee{-8}
 is an order of magnitude greater than in TMC-1, but is consistent with the
value found in L134N. More spatially sensitive observations are needed to
determine whether SO is enhanced in \candy.  

Early observations reported an enhancement of
\hcop\ towards IC~443 \citep{dickinson80,denoyer81}, which was
contrary to predictions for slow shocks \citep{iglesias78}, but
could be explained by the increased ionisation expected in
SNRs \citep{elitzur83}. However, subsequent observations found that the
\hcop\ abundance in IC~443 was not enhanced by the SNR shock
\citep{ziurys89} and even decreased in the high density gas, probably
due to more rapid dissociative recombination with electrons and 
reactions with H$_2$O \citep{vand93}. Our beam-averaged result 
 of 7.5\ee{-8} implies that the \hcop\ abundance is either enhanced in
\candy, or that that the source size used for the beam correction of the
\hcop\ data was incorrect. As for SO, more spatially-sensitive
observations are needed to determine if there is an enhanced abundance of
 \hcop\ in \candy.

\subsection{Shocked \h}
\label{sec-h2}

The measured \h\ 1--0/2--1 line ratio of 5--6 is somewhat lower
than the ratio of 10--20 expected for C-shocks \citep{burton88}. A
lower line ratio can be produced in J-shocks, where the
vibrational ground-states of \h\ are generated by a cascade following
collisional excitation of excited electronic states. However, this
type of shock cannot account for the high 1--0~S(1) line intensities
observed in \candy\ \citep{hol89}. A very low
line ratio ($\approx 2$) can be measured for fluorescent \h\ emission
produced by radiative excitation of low density gas by UV
photons \citep{black78}. UV excitation can also produce higher line ratios
in molecular gas of density \e{5}--\e{6}\cm{-3}
\citep*{burton90}. However, to generate 1--0~S(1) emission with a 
peak of 0.005\ergs, as found in \candy, the far-UV (FUV)
radiation field needs to be $>$ \e{5}~G$_0$, where
G$_0$ represents the FUV radiation field equivalent to
1.6\ee{-3}\ergs\ \citep{burton90}. This is an unusually high value to be
produced by only a few young stars. For example, early-type B stars
 typically produce a field of $\approx$ \e{3}~G$_0$ (note that strictly this is a 
 function of the distance from the star).  \citet{burton90} showed that
less than \e{-4} of the energy of  6--13.6~eV photons is
re-emitted in the 1--0~S(1) line. The total NIR \h\ 1--0 S(1) line 
luminosity of \candy\ is about 500~\lsol\ (Table~\ref{tab-h2}), 
so an exciting source would need to emit 5\ee{6}~\lsol\ in 6--13.6 eV photons, with
an even higher bolometric luminosity. Therefore, the \h\ emission we
detect is too bright and widespread to be produced by UV excitation
from a star and, despite the lower than expected line ratio, 
must be produced by the SNR shock. One possible explanation is  a 
 combination of gas components with different physical properties 
contributing to the detected \h\ emission. From observations of excited \h\ S(0)--S(7) lines 
with the {\em Spitzer Space Telescope} IRS spectrograph in 5--15 \micron\ and 
12--42.2 \micron\ bands, \citet{hewitt09} found that 
two components are required to produce the observed \h\ line intensity: a slow 
($\sim 10$\kms) C-type shock propagating into dense gas (\nh $\sim$ \e{6}\cm{-3}, 
$T\sim$500~K, $N$(\h)$\sim$2.8\ee{20}\cm{-2}) and a faster ($\sim 40$\kms) C-type 
shock propagating into lower density gas (\nh $\sim$ \e{4}\cm{-3}, $T\sim$1600~K, 
$N$(\h)$\sim$5.2\ee{18}\cm{-2}). They also detected ionic lines (mainly [Fe {\sc ii}]), 
which could be explained with models of a J-shock travelling in hot, low density gas 
(\nh $\sim 700$\cm{-3}, $T \sim 7000$~K).  Multiple gas components are often found in 
molecular clouds interacting with SNRs \citep[eg][]{vand93,reach05,hewitt06}. J-shocks 
will dissociate \h\ and then ionize the atomic hydrogen, eventually producing the \h\ 1--0 
 emission, while C-shocks will provide the \h\ 2--1 emission. Hence, a mix of shocks 
along the line of sight might produce both an elevated 1--0 emission for C-shocks and 
too high a  2--1 emission for the J-shock, resulting in the line ratio we observed. 

In shocked gas the 1--0~S(1) line contributes typically 5--10 $\%$ 
of the total \h\ luminosity \citep[e.g., see][]{burton92}. The \h\ luminosity of the source 
would therefore be high,  $\approx 5$\ee{3}\lsol\ , at the SNR distance of 18~kpc.  
This is approximately five times higher than the  $\approx$\e{3}\lsol\ emitted in \h\
lines by IC~443 \citep{burton88}. It is also brighter than 
the extended fluorescent \h\ line emission associated with 
several massive star-forming complexes \citep[e.g., 300\lsol\ from the
 PDR emission around the Orion Molecular Cloud-One;][]{burton90c}.

There is morphological evidence that the \h\ emission is related to
the SNR shock in \candy. Figure~\ref{fig-h2+20} shows that the \h\ filaments
coincide with the western edge of the bright continuum shell. 
The \h\ knots are associated with the locations of \oh masers, which are 
produced by collisional excitation. This morphological coincidence between  the \h\
and radio continuum emission, and between the \h\ and maser emission
suggests that these phenomena are shock-related. Furthermore, our \h\ emission is 
strikingly similar to emission detected by the {\em Spitzer Space Telescope} IRAC camera 
in the 5.8 band \citep{reach06}, which has a shocked origin, as confirmed 
by the IRS observations of \citet{hewitt09}. 

\citet{dubner04} have also mapped \candy\ in the \co\ 1--0, 2--1 and 3--2 transitions, 
with a somewhat different survey area.  While their integrated \co\ 1--0 and 2--1 maps are 
similar to our corresponding maps, their 3--2 map is distinct and matches the morphology 
of the \h\ emission well. The authors noted this difference and that their line widths and 
derived gas density do not imply shocked molecular gas. However, the higher than unity 
line ratio between the three \co\ transitions (1--0, 2--1 and 3--2) at the peak of 
the cloud  plus a possible asymmetry in the 
\co\ 2--1 and 3--2 line profiles (which we ascribe to self-absorption) are indicative of 
shocked gas. \citet{dubner04} suggest that the SNR is in front of the molecular cloud which
it is impacting, with the eastern side being pushed away from us, and the western side of the 
SNR being pushed towards us. 

The \h\ emission we detect shows a wide range of velocities from $-40$\kms to +40\kms. 
However, there are no molecular cloud components at $-40$ or +40\kms\ observed in 
our \co\  or OH data. This is consistent with similar observations obtained towards other 
SNRs associated with \oh masers G359.1--0.5 and G357.7--0.1 \citep{lazendic02,lazendic04}. 
It is likely that our 
arcsecond-scale NIR observations have the sensitivity to detect individually accelerated 
molecular clumps, while the mm and cm molecular data are sensitive to the bulk motion
of the molecular gas, with ambient velocities. The \h\ observations reveal the shocked 
gas viewed from different angles, while the mm observations mostly probe the molecular 
gas along the line-of-sight and the optically thicker and colder outside layers of the 
molecular cloud.  In the case of the SNR 3C~391, which is also associated with \oh 
maser emission, \citet{reach99} detected broad molecular lines ($\sim 20$\kms) 
only from a small gas clump (labeled 3C~391:BML) less than 0.6~pc in size. Such a 
feature would be about 6~arcsec in size if located at the distance of \candy. Comparing 
the distribution of NIR H$_2$ emission and \co\ emission for 3C~391, \citet{reach02} further 
showed that broad-line \co\ emission does follow shocked \h\ emission, but that the line 
strength of the broad-line component was 10 times weaker than for the 3C~391:BML 
clump, which coincides with the \oh maser. In summary, there are several factors which
could explain the non-detection of broad molecular lines in an SNR suspected of 
interaction with a molecular cloud: the lack of high spatial resolution in the observations, 
confusion with ambient gas (ie self-absorption) and with higher \co\ transitions 
\citep{reach99}. \citet{jiang2010} compiled a list of SNRs having confirmed or suspected 
interactions with molecular clouds, and of 34 SNRs confirmed (using various methods),
only 16 have broad and/or asymmetric molecular lines.

\subsection{Shocked OH and maser emission}

\begin{figure*} 
\centering
\includegraphics[height=10cm]{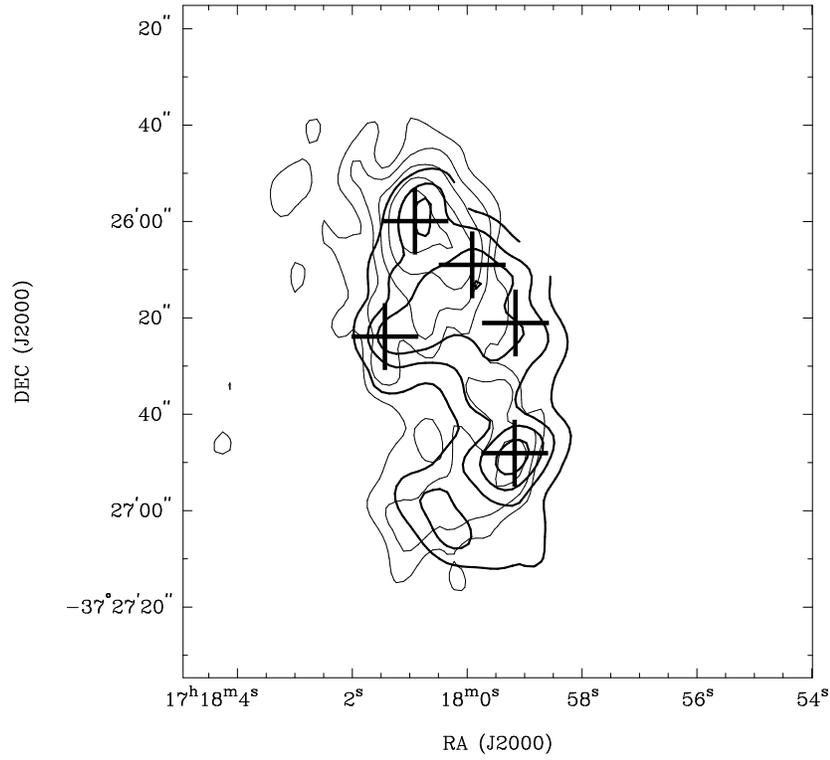}
\caption{Contours of OH(1667~MHz) absorption distribution between 10 and 20\kms\ (light lines) 
superimposed on that of \hone\ emission (heavy lines) which have been convolved to a
similar resolution. The two distributions show a good correlation. 
The OH contour levels are 
the same as in Figure~\ref{fig-oh+20}, and the \h\ contour
 levels are: 3.4, 10, 17 and 27 \ee{-4}\ergs . The crosses mark the \oh maser positions.}
\label{fig-oh+h2}
\end{figure*}

Figure~\ref{fig-oh+h2} shows contours of velocity-integrated
OH absorption at 1667~MHz overlaid on the contours of 
velocity-integrated \h\ emission. Good correlation between the OH and \h\ 
distributions is a strong
indication that the OH gas is generated in the shocked regions. 
 Other evidence to support the shock origin of the OH gas comes from the 
 association of the OH peaks with the masers. The OH column density derived in section $\S$3.7 
yields an OH fractional abundance of (0.4--1.0)\ee{-6}, indicating that 
 the OH abundance in \candy\ might be 3 times higher than in
 dark clouds \citep[e.g., $\sim$3\ee{-7};][]{ohishi92}.
An enhanced OH abundance towards an SNR (IC~443) was
first reported by \citet{denoyer79},  but shock models do not predict such
an enhancement \citep{draine83,hol89,kauf96}. For
temperatures above 400~K, any OH formed will be rapidly 
converted into H$_2$O through reactions with \h.  However,  X-rays
from the interior of the remnant can penetrate the surrounding
molecular cloud and eject photo-electrons, which then collisionally
excite the Werner Ly$\alpha$ band of \h, providing enough energy to
dissociate the H$_2$O molecules into OH, but not enough to dissociate the OH
\citep{wardle99}. The process of forming abundant OH occurs when the temperature 
drops to a point where the conversion of OH back into H$_2$O ceases. A warm, dense
layer rich in OH develops at the rear of the shock, ideal for the
collisional pumping of  \oh masers.  The upper value of the derived OH
abundance in \candy\ is consistent with predictions for 
the dissociation of $1\%$ of the water by the SNR X-ray flux  \citep{wardle99}.

Our non-detection of excited-state OH masers is consistent with previous searches: \citet{fish07} searched for 6~GHz OH masers towards 10 SNRs with the Effelsberg 100-m telescope; \citet{pihlstrom08} searched for 4.7, 7.8, 8.2 and 23.8~GHz OH transitions  in 
4 SNRs with the Very Large Telescope; and  \citet{mcdonnell08} 
searched for the 6~GHz OH transitions with the Parkes 64-m telescope in about 40 SNRs, 
including sources from the Magellanic clouds and \candy.  These results suggest that 
the shock conditions suitable for the production of 1720~MHz OH transitions do not 
support the other OH transitions, even those at 6049~MHz and 4765~MHz, which  have
the most compatible requirements. Interestingly, 
\citet{wardle07}  found that for OH column densities larger than 
\e{17}\cm{-2}, the 6049~MHz masers switch on while the 1720~MHz transition is quenched,
even with suitable gas densities and temperatures for the production of 1720~MHz masers,
 suggesting that the higher frequency  line might 
serve as a complementary signpost of warm, shocked gas. Unfortunately, a dedicated 
search for the 6049~MHz OH line by \citet{mcdonnell08} was unsuccessful.  However, 
they found indications that this transition is more sensitive than the 1720~MHz line to 
velocity coherence. Similarly, searches for the 22~GHz H$_2$O 
maser transition in about 20 SNRs reported no detections \citep{claussen99,woodall07}. 
Detailed modeling of this transition using the conditions required for 
1720~MHz OH maser production with both C-type and J-type shocks indicate that 
by the time the appropriate gas density is reached ($\sim$\e{9}\cm{-3}), the gas 
temperature has cooled too much (to 15~K from required 300-600~K) to be able to 
generate the  necessary H$_2$O column density \citep{woodall07}.

\section{Conclusion}
\label{sec-candy5}

Radio and infrared observations towards \candy, an SNR with associated  \oh masers, 
were used to investigate the interaction of the SNR with the adjacent 
molecular cloud. The main results are  summarised below.

\begin{enumerate} 
\renewcommand{\theenumi}{(\arabic{enumi})}

\item{Emission from several molecular species (\co, \tco, CS,
\hcop, HCN, \htco\ and SO) have been detected towards the SNR at
the \oh maser velocities.  The molecular lines have moderate line widths 
($\approx 4$\kms), showing no kinematic evidence of shock, but the derived gas
 density of $\approx$\e{5}\cm{-3} and temperature of $\approx
 45$~K are consistent with the predictions for \oh maser production in 
molecular gas triggered by an SNR shock.} 

\item{Molecular abundances of the shocked molecular gas in  \candy\ are
found to be somewhat different from those in another well-studied SNR, 
IC~443. The abundances of molecules such as CS, HCN and \htco\  are
 not enhanced by the effect of the shock, as expected from the most
plausible models. The abundances of \hcop\ and SO appear to be an order of
magnitude higher than in dark clouds. 
No SiO emission was detected, but the upper limit to
the SiO column density does allow the possibility of an 
enhanced SiO abundance.} 

\item{Strong \h\ emission was detected towards the SNR. Its distribution
coincides with the bright radio shell and the location of the OH
masers, supporting the proposal that the \h\ emission arises from the
SNR shock expanding into the adjacent molecular cloud.
The total \h\ line luminosity in the source is very 
large, $\approx 5$\ee{3}~\lsol, which indicates that the interaction of SNRs 
with molecular clouds can produce \h\ line emission greater
than that associated with fluorescent emission around
massive star-forming complexes.  This should be born in mind when 
interpreting extragalactic \h\ line emission.}

\item{In \candy\ we have clear observational evidence
of an extended OH cloud associated with shocked regions in the SNR, as 
revealed by thermal absorption against the SNR continuum emission. The
distribution of the OH gas correlates well with the
\oh maser locations and the distribution of shocked \h. An 
 upper value for the OH fractional abundance ($\sim$\e{-6}) is greater than for
 cold clouds and is consistent with predictions 
for water dissociation by a soft X-ray flux from the SNR. We found no
\htco\ absorption towards the SNR.} 

\item{We detected for the first time OH main-line maser
emission at 1665 and 1667~MHz in directions towards and near \candy. 
Only a 1665~MHz maser  is located towards the SNR and was identified with
an IR source SSTGLMC~G349.7294+00.1747 from the GLIMPSE Catalog. The 
other maser was detected in both main-line transitions and coincides with  the ultra-compact H\,{\sc ii}
region IRAS 17147--3725, located south-east of the SNR. We also 
found weak \htco\ absorption towards this \hr\ region.
No masing was detected towards the SNR from excited-state OH transitions 
around 4.8 and 6~GHz, nor from the 6~GHz CH$_3$OH transition. }

\end{enumerate}

\section*{Acknowledgments}

We thank John Black for kindly providing us with his RADEX code, 
Estela Reynoso for use of the 12-m NRAO CO map, and
Andrew Walsh for help with the CASPIR observations. We would also like to thank our referee, John Phillips, who's comments have greatly 
improved this manuscipt. JSL and JBW acknowledge
travel support from the Australian Government's Access 
to Major Research Facilities Program (AMRFP).  This work was also
supported in part by the Australian Research Council and Australian's Government International Postgraduate Research Scholarship program. 

The Australia Telescope Compact Array and Mopra Telescope are 
part of the Australia Telescope  funded by the Commonwealth 
of Australia for operation as a National Facility, managed by CSIRO. 
The Swedish-ESO Submillimetre Telescope (SEST) is operated by the
Swedish National Facility for Radio Astronomy, Onsala Space
Observatory and by the European Southern Observatory (ESO).
The NIR observations would not have been 
possible without the efforts of Michael Ashley and the UNSWIRF crew from
UNSW, as well as the staff of the Anglo Australian Observatory.
CASPIR is operated by the  Australian National University  at the
2.3 m telescope at Siding Spring Observatory. 
This publication makes use of data products from the Two Micron All Sky Survey, which is a joint project of the University of Massachusetts and the Infrared Processing and Analysis Center/California Institute of Technology, funded by the National Aeronautics and Space Administration and the National Science Foundation. This research has made use of the NASA/ IPAC Infrared Science Archive, which is operated by the Jet Propulsion Laboratory, California Institute of Technology, under contract with the National Aeronautics and Space Administration.

%%%%%%%%%%%%%%%%%%%%%%%%%%%%%%%%%%%%%
\clearpage

\begin{table*}
\centering
\caption{The \oh masers detected towards the SNR \candy\  (from \citealt{frail96}). Given are maser designations, as used in this paper, positions, peak flux density (S$_p$), peak velocity (V$_{\rm {LSR}}$) and velocity resolution ($\Delta v$).}
\begin{tabular}{@{}cccccc}
\hline\hline
Designation &  RA(2000) & Dec.(2000) & S$_p$ & V$_{\rm {LSR}}$ & $\Delta v$ \\ 
   & (h m s) & (\degr\  \arcmin\  \arcsec\ ) & (mJy) & (\kms) &  (\kms)\\
\hline
M1 & 17 17 59.2 & -37 26 21.07 & 90 & $+$16.7 & 1.5 \\ 
M2 & 17 17 59.2 & -37 26 48.07 & 152 & $+14.3$ & 1.8 \\
M3 & 17 17 59.9 & -37 26 09.01 & 1310 & $+$16.0 & 1.8 \\
M4 & 17 18 00.9 & -37 25 59.94 & 1020 & $+$15.2 & 1.6 \\
M5 & 17 18 01.4 & -37 26 23.90 & 277 & $+$16.9 & 1.4 \\
\hline
\end{tabular}
\label{tab-masers}
\end{table*}

%%%%%%%%%%%%%%%%%%%%%%%%%%%%%%%%%%

\begin{table*}
\centering
\caption{Mopra and SEST observational parameters for
\candy.}
\begin{tabular}{@{}lcc}
\hline\hline
Parameter & Mopra & SEST \\
\hline
Observed frequencies (GHz) &  110 &  86 -- 270 \\
Total bandwidth (MHz)      &   64 & 2$\times$64 \\
No. of frequency channels  & 1024 & 2$\times$1000 \\
Velocity resolution (km\,s$^{-1}$) & 0.20 & 0.06 -- 0.14 \\
Velocity coverage (km\,s$^{-1}$)   & 160 & 60 -- 100 \\
FWHP beamwidth (arcsec) & 43 & 57--20 \\
Area observed (arcmin$^{2}$)    & 3.5 $\times$ 4 & 1.5 $\times$ 2\\
Grid interval (arcsec)    & 30 &  24 \\ 
Integration time (min) & 1 & 1 \\
\hline
\end{tabular}
\label{tab-molecules}
\end{table*}

%%%%%%%%%%%%%%%%%%%%%%%%%%%%%%%%%%

\begin{table*}
\centering
\caption{ATCA observational parameters for \candy.}
\begin{tabular}{@{}lcccccc}
\hline\hline
Parameter & \multicolumn{4}{c}{OH observations} & \htco & CH$_3$OH \\
\hline
Date  & Dec 1999 & Nov 1998 & Nov 1998 & May 2001  & May 2000 & Apr 2002\\
      & Jun 2000 &          &          &           &    \\
Central frequency (MHz) & 1666 & 4660 & 6030 & 6049  & 4829 & 6668 \\
 &      & 4751 & 6035 &       &  \\ 
 &      & 4765 &      &       &  \\
Array configurations   & 1.5A, 6B & 6D & 6D & 750A & 1.5A & 6A \\
Primary beam (arcmin)  & $\approx$ 33 & $\approx$ 10 & $\approx$ 10 &
$\approx$ 10 & $\approx$ 10  & $\approx$ 10 \\
Total bandwidth (MHz): & \\
~spectral-line & 8 & 4 & 4 & 4  &  4 & 4  \\
~continuum & 128 & 128 & 128 & --   &  -- & --  \\
No. of frequency channels: & \\
~spectral-line & 1024 & 1024  & 1024 & 2048 &  2048 &  2048\\
~continuum  & 32 & 32  & 32 & -- &  -- &  -- \\
Velocity resolution (km\,s$^{-1}$) & 1.6 & 0.2 & 0.2 & 0.1 & 0.1 & 0.1  \\
Total observing time (hr) & 2$\times$13 & 13 & 13 & 3  & 13 & 6 \\
Spectral-line cubes:     &             &    &    &    &   & \\
~~Synthesised beam (arcsec$^2$) & 15 $\times$ 12 & 3 $\times$ 2 &
2 $\times$ 2 & 37 $\times$ 10 & 4 $\times$  2 & 2 $\times$ 1.5 \\
~~rms noise (\mjb)  & 5  & 7 & 6 & 9  & 7 & 9 \\
\hline
\end{tabular}
\label{tab-atca}
\end{table*}

%%%%%%%%%%%%%%%%%%%%%%%%%%%%%%%%%%

\begin{table*}
\centering
\caption{Molecular transitions observed with ATCA towards \candy.}
\begin{tabular}{@{}lll}
\hline\hline
Molecule & Frequency & Transition  \\
\hline

OH & 1665 MHz & $^2\Pi_{3/2}$ J=3/2 \\
OH & 1667 MHz & $^2\Pi_{3/2}$ J=3/2 \\
OH & 4660 MHz & $^2\Pi_{1/2}$ J=1/2 \\
OH & 4751 MHz & $^2\Pi_{1/2}$ J=1/2 \\
OH & 4765 MHz & $^2\Pi_{1/2}$ J=1/2 \\
OH & 6031 MHz & $^2\Pi_{3/2}$ J=5/2 \\
OH & 6035 MHz & $^2\Pi_{3/2}$ J=5/2 \\
OH & 6049 MHz & $^2\Pi_{3/2}$ J=5/2 \\
\htco\ & 4829 MHz & $1_{11} - 1_{10}$ \\
CH$_3$OH & 6668 MHz & $5_1 - 6_0$A+\\

\hline
\end{tabular}
\label{tab-atca-trans}
\end{table*}

%%%%%%%%%%%%%%%%%%%%%%%%%%%%%%%%%%%%%%

\begin{table*}
\begin{center}
\caption{The first three columns list the observed molecular
transitions and their frequencies ($\nu$) towards the SEST peak
position, except for the CS and \htco\
transitions which are obtained towards M1. 
The fourth column lists the SEST and Mopra FWHM beam size, 
the fifth gives the observed main-beam brightness temperature (\tmb) and
rms noise (upper limits are 2$\sigma$ values), while the last two
columns list line velocities ($v_{\rm LSR}$) and FWHM line widths
($\Delta v_{\rm LSR}$).\label{tab-candy-sest}}
\begin{tabular}{@{}lcccccc}
\hline \hline
Molecule & Transition &  $\nu$~~ & Beam Size & $T_{mb}$ & $v_{\rm LSR}$ &
$\Delta v_{LSR}$  \\
  &  & (GHz) & (arcsec) &  (K)  & (\kms) & (\kms) \\ 
\hline
\tco$^a$ & 1--0 & 110.197 & 43 &  3.9$\pm$0.2 & 16.4 & 2.7 \\
\co   & 2--1 & 230.538 & 23 & 31.1$\pm$0.1 & 15.1 & 3.9   \\
       & 1--0 & 115.271 & 45 & 17.8$\pm$0.2 & 16.5 & 3.9 \\
CS    & 5--4 & 244.935 & 21 & $<$0.4 & --   &  --  \\
       & 3--2 & 146.969 & 34 & 0.7$\pm$0.1& 16.1 & 4.2 \\
       & 2--1 & 97.981  & 52 & 0.5$\pm$0.1 & 16.7 & 3.3  \\
\hcop & 3--2 & 267.557 & 20 & $<$0.3 & --   &   --  \\
    & 1--0 & 89.188  & 54 & 2.1$\pm$0.1 & 13.3 & 4.5  \\
HCN   & 3--2 & 265.886 & 20 & $<$0.5 & --  &   --  \\
  & 1--0$^b$ & 88.632  & 55 & 1.0$\pm$0.1 & 12.9 & 3.3 \\
 \htco & 3$_{(1,2)}$--2$_{(1,1)}$ & 225.698 & 23 & $<$0.2 & -- & -- \\ 
     & 3$_{(0,3)}$--2$_{(0,2)}$ & 218.222 & 24 & $<$0.2 & -- & -- \\
     & 2$_{(1,1)}$--1$_{(1,0)}$ & 150.498 & 36 & 0.3$\pm$0.1 & 16.8 & 6.5 \\
     & 2$_{(1,2)}$--1$_{(1,1)}$ & 140.839 & 38 & $<$0.2 & -- & -- \\
SO   & 2$_3$--1$_2$ & 109.252 & 47 & $<$0.2 & --   & --   \\
    & 3$_2$--2$_1$ &  99.299 & 50 & 0.5$\pm$0.1 & 14.7 & 3.0  \\
SiO  & 5--4 v=0 & 217.105 & 24 & $<$0.4 & --   & --   \\
     & 3--2 v=0 & 130.268 & 40 & $<$0.3 & --   & --     \\
     & 2--1 v=0 &  86.846 & 57 & $<$0.2 &  --  & --    \\
\hline
\end{tabular}
\end{center}
($a$) from Mopra observations; (b) values for the main hyperfine component;
\end{table*}

%%%%%%%%%%%%%%%%%%%%%%%%%%%%%%%%%%

\begin{table*}
\centering
\caption{Summary of results from UNSWIRF observations. The knots
denote \h\ peaks near the five \oh masers. All sets of features are 
numbered correspondingly. The second and third columns
list the peak flux density of 1--0 and 2--1 S(1) \h\ lines, corrected
for extinction as follows: $A_{2.12}$ = 3\,mag, $A_{2.25}$ =
2.7\,mag. The fourth and fifth  columns list the total flux density of corresponding \h\ lines, 
corrected for extinction. The sixth and seventh columns list
the luminosity of the corresponding \h\ lines for a source distance of 18~kpc. The eighth 
column gives the line ratio R = L(1--0)/L(2--1). The absolute flux density measurements
 have less than 30\% uncertainty.}
\begin{tabular}{@{}lccccccc}
\hline\hline
Designation & $F_p$(1--0) &  $F_p$(2--1) & $F_{int}$(1--0) &
$F_{int}$(2--1) &     L(1--0) & L(2--1) & R \\
  & \multicolumn{2}{c}{(\ee{-4}\ergs)} & \multicolumn{2}{c}{(\ee{-13}~ergs\,s$^{-1}$\,cm$^{-2}$)} & (\lsol) & (\lsol)  &  \\ 
\hline
Knot 1 & 24 & 6.0 & 38 & 8.4 & 38 & 8 & 5  \\
Knot 2 & 52 & 9.6 & 66 & 12 & 66 & 12 & 5 \\
Knot 3 & 43 & 8.4 & 49 & 10 & 49 & 10 & 5 \\
Knot 4 & 36 & 7.2 & 40 & 4.8 & 40 & 5 & 8  \\
Knot 5 & 44 & 8.4 & 41  & 7.2 & 41 & 7 & 6  \\
total  & 52 & 9.6 & 540 & 90 & 540 & 88 & 6 \\
\hline
\end{tabular}
\label{tab-h2}
\end{table*}

%%%%%%%%%%%%%%%%%%%%%%%%%%%%%%%%%%

\begin{table*}
\centering
\caption{Parameters of the main-line OH masers at 1665 and 1667~MHz 
detected towards the region of \candy.}
\begin{tabular}{@{}lcc}
\hline\hline
Parameter & MM(1) & MM(2) \\
\hline
RA (2000) (h m s) & 17 17 59.3 & 17 18 11.1 \\
Dec. (2000) (\degr~\arcmin~\arcsec~) & $-$37 26 10.00 & $-$37 28 23.93 \\
1665 peak (Jy)               & 0.09 & 0.06 \\
1665 V$_{\rm {peak}}$ (\kms) & 21   & 16 \\
1667 peak (Jy)               & --   & 0.10 \\
1667 V$_{\rm {peak}}$ (\kms) & --   & 10 \\
\hline
\end{tabular}
\label{tab-mainl}
\end{table*}

%%%%%%%%%%%%%%%%%%%%%%%%%%%%%%%%%%

\begin{table*}
\caption{Molecular column densities, $N$(mol), and abundances, 
$X$=$N$(mol)/$N$(\h) for $N$(\h)=4\ee{21}\cm{-2}, of the molecular 
cloud interacting with the SNR \candy. For comparison, abundances of
the shocked molecular gas in SNR IC~443 \citep{vand93} are 
listed, as well as those of the dark clouds TMC-1
 and L134N \citep{ohishi92}. Note that abundance values are always 
assumed for \co\ and \tco. \label{tab-abundances}}
\begin{tabular}{@{}lrrrrrr}
\hline\hline
Molecule & \multicolumn{2}{c}{\candy} & \multicolumn{2}{c}{IC 443$^a$}
& TMC-1 & L134N \\
  & $N$ (\cm{-2}) &   $X$  & $X_1$ & $X_2$    & $X$ & $X$  \\
\hline
\co   & 4\ee{17}    & 1\ee{-4}     & \e{-4}      & \e{-4}    & 8\ee{-5}  & 8\ee{-5} \\
\tco  & 8\ee{15}    & 2\ee{-6}     & 2\ee{-6}    & 2\ee{-6}  & ... & ...  \\
\hcop & 3\ee{14}    & 7.5\ee{-8}   & 1\ee{-8}    & 3\ee{-10} & 8\ee{-9} & 8\ee{-9}\\
HCN   & 1.3\ee{14}  & 3.2\ee{-8}   & 3\ee{-8}    & 9\ee{-9}  & 2\ee{-8} & 4\ee{-9} \\
CS    & 3\ee{13}    & 7.5\ee{-9}   & 6\ee{-9}    & 8\ee{-9}  & \e{-8} & \e{-9}\\
SO    & 1.2\ee{14}  & 3.0\ee{-8}     & \multicolumn{2}{c}{8\ee{-9}$^b$} & 5\ee{-9} & 2\ee{-8}\\
\htco & $<$5\ee{13} & $< 1.2$\ee{-8}   & 7\ee{-9}    & 2\ee{-8}  & 2\ee{-8} & 2\ee{-8} \\
SiO   & $<$3\ee{13} & $< 7.5$\ee{-9}  & $<$2\ee{-9} & 4\ee{-9}  &
$<$2\ee{-12} & $<$4\ee{-12}\\

\hline
\end{tabular}

\medskip
({\em a})  It is suggested that the shocked gas in IC 443 has
two components: $X_1$ abundances correspond to shocked gas with
\nh $\sim$ \e{5}\cm{-3} and \tkin\ $\sim$ 80 K, while $X_2$ abundances
correspond to shocked gas with \nh $\sim$ 3\ee{6}\cm{-3} and \tkin\
$\sim$ 200~K \citep[see][for more details]{vand93}; ({\em b}) derived for a
single component model; 
\end{table*}

%%%%%%%%%%%%%%%%%%%%%%%%%%%%%%%%%%
%%%%%%%%%%%%%%%%%%%%%%%%%%%%%%%%%%

\clearpage

%1

%%%%%%%%%%%%%%%%%%%%%%%%%%%%%%%%%%
\clearpage
%2

%%%%%%%%%%%%%%%%%%%%%%%%%%%%%%%%%%

\clearpage
%3

%%%%%%%%%%%%%%%%%%%%%%%%%%%%%%%%%%

%\clearpage
%4

%%%%%%%%%%%%%%%%%%%%%%%%%%%%%%%%%%%%
\clearpage
%45 insert

%%%%%%%%%%%%%%%%%%%%%%%%%%%%%%%%%%%%
\clearpage
%5 (6)

%%%%%%%%%%%%%%%%%%%%%%%%%%%%%%%%%%
\clearpage
%6 (7)

%%%%%%%%%%%%%%%%%%%%%%%%%%%%%%%%%%
%\clearpage
%7 (8)

%%%%%%%%%%%%%%%%%%%%%%%%%%%%%%%%%%
\clearpage
%8 (9)

\clearpage
%9 (10)

%%%%%%%%%%%%%%%%%%%%%%%%%%%%%%%%%%

\clearpage
%10 (11)

%%%%%%%%%%%%%%%%%%%%%%%%%%%%%%%%%%
\clearpage
%11 (12)

%%%%%%%%%%%%%%%%%%%%%%%%%%%%%%%%%%
\clearpage
%12 (13)

%%%%%%%%%%%%%%%%%%%%%%%%%%%%%%%%%%

%%%%%%%%%%%%%%%%%%%%%%%%%%%%%%%%%%

%\label{lastpage}

\end{document}